\documentclass{aa}  
\usepackage{longtable}
\usepackage{caption}
\usepackage{subcaption}
\usepackage{amsmath}
\usepackage{graphicx}
\usepackage{txfonts}
\usepackage[fleqn]{amsmath}	% Advanced maths commands
\usepackage{amssymb}	% Extra maths symbols
\usepackage{mathtools}
\usepackage[breaklinks=true]{hyperref}
\hypersetup{
  colorlinks=true,   %% links colored instead of frames
  citecolor=blue,    %% links' color
  urlcolor=blue,     %% external hyperlinks
  linkcolor=blue,     %% internal latex links (eg Fig)
}
\bibpunct{(}{)}{;}{a}{}{,} %% natbib format for A&A and ApJ
\usepackage{lscape}

\DeclareRobustCommand{\ion}[2]{%
\relax\ifmmode
\ifx\testbx\f@series
{\mathbf{#1\,\mathsc{#2}}}\else
{\mathrm{#1\,\mathsc{#2}}}\fi
\else\textup{#1\,{\mdseries\textsc{#2}}}%
\fi}

\newcommand{\Feii}{\ion{Fe}{ii}\,}
\newcommand{\Hb}{\ensuremath{\mathrm{H}\beta}\xspace} % THis works in both text and equations.
\newcommand{\Rfe}{$R_{\rm Fe}$\,}
\usepackage{xspace} % This package adds a space when needed. You can remove the "\," at the end of \Rfe, \Hb, etc and replace with \xspace.
\newcommand{\ergs}{\ensuremath{\mathrm{erg\,s^{-1}}}\xspace}
\let\oldAA\AA
\renewcommand{\AA}{\text{\oldAA}\xspace}
\newcommand{\jwst}{\textit{JWST}\xspace}

\begin{document}

    \title{The missing \ion{Fe}{ii} bump in faint \textit{JWST} AGN: possible evidence for metal-poor broad-line regions at early cosmic times}

   \author{Bartolomeo Trefoloni\inst{1,2,3}
   \thanks{\email{bartolomeo.trefoloni@unifi.it}},
    Xihan~Ji\inst{3,4},
    Roberto~Maiolino\inst{3,4,5}, 
    Francesco~D'Eugenio\inst{3,4},
    Hannah~\"Ubler\inst{3,4,6},
    Jan~Scholtz\inst{3,4},
    Alessandro~Marconi\inst{1,2},
    Cosimo~Marconcini\inst{1,2},
    Giovanni~Mazzolari\inst{7}
	      }
          
% List of institutions
\institute{
$^{1}$Dipartimento di Fisica e Astronomia, Universit\`a di Firenze, via G. Sansone 1, 50019 Sesto Fiorentino, Firenze, Italy\\
$^{2}$INAF -- Osservatorio Astrofisico di Arcetri, Largo Enrico Fermi 5, I-50125 Firenze, Italy\\
$^{3}$Kavli Institute for Cosmology, University of Cambridge, Madingley Road, Cambridge, CB3 OHA, UK\\
$^{4}$Cavendish Laboratory - Astrophysics Group, University of Cambridge, 19 JJ Thomson Avenue, Cambridge, CB3 OHE, UK\\
$^{5}$Department of Physics and Astronomy, University College London, Gower Street, London WC1E 6BT, UK\\
$^{6}$Max-Planck-Institut f{\"u}r Extraterrestrische Physik (MPE), Gie{\ss}enbachstra{\ss}e 1, 85748 Garching, Germany\\
$^{7}$INAF -- Osservatorio di Astrofisica e Scienza dello Spazio di Bologna, via Gobetti 93/3, I-40129 Bologna, Italy\\
\\
}

\titlerunning{The missing \ion{Fe}{ii} bump in faint \textit{JWST} AGN}
\authorrunning{B. Trefoloni et al.}

\abstract{Recent \textit{JWST} observations have revealed a large population of intermediate/low-luminosity active galactic nuclei (AGN) at early times with peculiar properties, different from local AGN or luminous quasars. To better understand the physical conditions in the broad-line regions (BLRs) of these early AGN, we used the optical \Feii (4434--4684 \AA) and the broad \Hb emission, and the ratio between their equivalent widths \Rfe, as a probe on a purposefully assembled sample. Specifically, we gathered a sample of 26 high redshift ($\langle z \rangle$=6.4) AGN, observed by \textit{JWST}, with broad \Hb detection both in the high and low luminosity regimes (respectively 14 faint AGN and 12 quasars), to investigate their optical \Feii emission properties. In addition, we carefully selected control samples at lower $z$. We found that the population of faint AGN ($\rm \log(L_{\Hb} / (erg \, s^{-1}))\lesssim 44$) exhibits a significantly lower \Feii emission than their local counterparts (\Rfe$<$0.24 versus \Rfe$\simeq$0.85 in the control sample), while the quasars at the epoch of reionisation observed by \textit{JWST} present a \Feii emission profile that closely resembles that observed at $z<3$. We argue that the weakness of the \Feii bump in the faint \textit{JWST} AGN might be due to the reduced metallicity of their broad line region ($\lesssim 0.5~Z_{\odot}$), while luminous quasars have already reached chemical maturity ($\sim Z{_\odot}$ or higher). Lastly, we highlight an intriguing similarity between the spectral properties of the high redshift population of faint AGN with those harboured in local metal poor dwarf galaxies.}

   \keywords{quasars: general -- quasars: supermassive black holes -- quasars: emission lines -- Galaxies: active --  Galaxies: Seyfert -- Galaxies: high-redshift }

   \maketitle
%
%-------------------------------------------------------------------

\section{Introduction}

Active Galactic Nuclei (AGN) are unanimously considered to be key agents in the process of shaping galaxies. The energy released from gas accretion onto super-massive black holes (SMBHs) powering the AGN has been shown capable of significantly affecting the star-formation processes within their host galaxies by heating and/or depleting the interstellar medium (ISM, e.g. \citealt{king2005agn, fabian2012observational, costa2015fast, king2015powerful}). As a consequence, repeatedly injecting energy into the surrounding ISM, AGN can ultimately lead to the quenching of the galactic star formation.
Tracking their ubiquity and investigating their properties through the cosmic ages enables us to follow the assembly history of the Universe.

Recently, our knowledge of the high-redshift ($z\sim 5-11$) Universe has been dramatically expanded through the observations obtained with the \textit{James Webb Space Telescope} (\textit{JWST}; \citealt{gardner2006james}). Several deep surveys (down to $mag_{F444W}=30.65$) carried out within the first \textit{JWST} cycles have indeed revealed a large population of intermediate/low bolometric luminosity ($\rm L_{bol} \sim 10^{42} - 10^{45} \, erg \, s^{-1}$) AGN with black hole masses ($M_{BH}$) already grown up to $10^{6}-10^{8} \, M_{\odot}$ within $z \sim$ 6-8 (\citealt{onoue2020no, kocevski2023hidden, harikane2023jwst, maiolino2023jades, larson2023ceers, ubler2023ga, greene2024uncover, ubler2024ga}) undetected by the previous optical and X-ray surveys of high-z quasars (see e.g. \citealt{yang2023desi} and references therein). 

% Many of these objects are routinely identified because of the segregation in the colour-colour space (red optical and blue UV colors) as well as for their morphological compactness (\citealt{labbe2023population, barro2023extremely, matthee2024little, hainline2024investigation}). In most of the cases, spectroscopic follow-ups have confirmed their nature of broad line AGN, with a detection efficiency of about $60\%$ (\citealt{greene2024uncover}, see \citealt{kokubo2024challenging} for other open issues). Additionally, new emission-line based diagnostics are being developed to efficiently identify the submerged population of Type II AGN (\citealt{mazzolari2024new} and references therein).

Most of these new AGN at high-$z$ are identified through the detection of a broad component of H$\alpha$ or H$\beta$ without a counterpart in [\ion{O}{iii}], hence excluding an outflow scenario and leaving the Broad Line Region (BLR) around an accreting black hole as the main plausible explanation (\citealt{kocevski2023hidden, ubler2023ga, maiolino2023jades, matthee2024little, kocevski2024rise, ubler2024ga, kokorev2023uncover, greene2024uncover, taylor2024broad}), although some more exotic scenarios have been proposed (\citealt{kokubo2024challenging, baggen2024small}). Type 2, narrow line AGN have also been searched, although with the caveat that standard BPT diagrams seem to lose their capability of discriminating between star forming galaxies and AGN at such early epochs (likely because of the low metallicity, \citealt{ubler2023ga, maiolino2023jades}), prompting the exploration of other Narrow Line diagnostics (\citealt{scholtz2023jades, chisholm2024ne, mazzolari2024new}). Interestingly, a number of these newly discovered AGN have peculiar colors, with red optical slopes and blue UV slopes (\citealt{labbe2023population, barro2023extremely, greene2024uncover, kocevski2024rise}) and have been dubbed Little Red Dots, although they contribute to only 10--30\% of the population of newly discovered AGN (\citealt{maiolino2023jades, hainline2024investigation, kocevski2024rise})

Interestingly, the photometric colours are not the only peculiarities observed in these sources. For instance, their black holes appear to be overmassive with respect to the stellar mass contained within the host galaxy, when compared to the expectations from scaling relations in the local Universe (\citealt{maiolino2023jades, ubler2023ga, harikane2023jwst, kokorev2023uncover, furtak2024high, juodvzbalis2024dormant, parlanti2024ga, marshall2024ga}). Such high-redshift overmassive BHs are predicted by several theoretical models as a direct consequence of super-Eddington accretion and/or direct-collapse black holes (\citealt{trinca2022low, koudmani2022two, schneider2023we}).

Another noticeable feature of faint \textit{JWST} AGN is their X-ray weakness. These sources are systematically undetected even in the X-ray stacks of the deepest \textit{Chandra} fields, such as GOODS-N and GOODS-S (\citealt{maiolino2024jwst,yue2024stacking,  kocevski2024rise, wang2024rubies}). Yet, it is still a matter of debate whether the observed X-ray weakness is actually intrinsic (i.e. due to an inefficient or beamed coronal emission, expected in some models, \citealt{pacucci2024mildly, madau2024x, maiolino2024jwst, king2024black} ) or rather caused by Compton-thick absorption along the line of sight (for a more detailed discussion see \citealt{maiolino2024jwst}).

In addition, low-luminosity \textit{JWST} AGN seem to lack the signature of prominent large-scale ionised winds, which are instead observed even in the low-luminosity tail of the local AGN population (\citealt{shenho2014, bisogni2017inclination}). This could be explained, at least qualitatively, by considering that the low gas metallicity observed in the narrow line region (NLR) of these sources implies lower dust content and consequently lower radiation pressure powering the outflow (\citealt{maiolino2024jwst}).

However, a physical picture embracing all the peculiarities featured by this new, elusive population is still far from being formulated. In this framework, valuable pieces of information can be gathered by a careful comparison with typical local AGN, matched in terms of accretion parameters (i.e. luminosity and black hole mass).

In the more local Universe ($z<1$), AGN are generally observed to share an ensemble of correlations between spectral properties which define the so-called Eigenvector 1, firstly discovered on a sample of 80 Palomar-Green AGN by \citet{boroson1992emission}. Several following studies aimed at consolidating the observational trends on more sound statistical basis, and arranged the spectral diversity of local AGN into a four-dimensional correlation space, the so-called 4DE1 (e.g. \citealt{sulentic2000phenomenology, zamfir2010detailed, marzianisulentic2014}, see also \citealt{marziani2018main} for a more comprehensive review). These properties include, among the other features, the anti-correlation between the strength of the narrow [\ion{O}{iii}] and of the \Feii (e.g. \citealt{shenho2014}), the anti-correlation between the full width at half maximum (FWHM) of the \Hb emission line and the ratio between the equivalent width (EW) of the \Hb and that of the \Feii (\Rfe; see e.g. \citealt{deconto2023high} and references therein), and the anti-correlation between the modulus of the \ion{C}{iv} emission line offset and its EW (e.g., \citealt{richards2011, rivera2022exploring, stepney2023no}). Additionally, other noticeable correlations have been observed across different wavebands, with the strength of the \Feii anti-correlating with both the Radio intensity (\citealt{miley1979relations}) and compactness (\citealt{osterbrock1977spectrophotometry}). In a similar fashion, generally steeper X-ray spectra (i.e. larger photon indices $\Gamma$) are found in objects with stronger \Rfe (\citealt{wang1996, laor1997soft, shenho2014}).

Although the ultimate physical driver(s) of the 4DE1 has not been fully understood, it is generally believed that the Eddington ratio and the inclination of the accretion disc along the line of sight play a crucial role in producing the observed diversity in terms of spectral shapes (\citealt{shenho2014, sun2015dissecting}, but see also \citealt{panda2018modeling} for a discussion on the effect of the Eddington ratio).  Also, a more thorough understanding of the mechanism underlying the optical 4DE1 trends is hampered by a meagre comprehension of the physical details of the \Feii emission, whose modelling is still not quite satisfactory. This is mostly due to the the complexity of a detailed treatment of the the \Feii ion and the fact that an accurate set of radiative and collisional atomic data is necessary to deal with the selective excitation, the continuum pumping and the fluorescence, which are relevant for the \Feii (see e.g. \citealt{sarkar2021improved} and references therein). Additionally, the physical mechanism responsible for the micro-turbulence (i.e. the effective turbulent motions within the line-forming region of the cloud), which is required to reproduce the strength of \Feii emission in observations (\citealt{netzer1983broad, baldwin2004origin, bruhweiler2008modeling}), is still far from being understood.

Although a theoretical framework explaining all the 4DE1 details has not been developed yet, this observational parameter space offers the possibility to track systematic differences and similarities between low- and high-redshift objects in a common (and model independent) parameter space.

In this work we aim at characterising the strength of the optical \Feii bump between 4434--4684 \AA, whose intensity is commonly included among the 4DE1 parameters. In particular we show that, despite faint \textit{JWST} AGN share the locus occupied by some of their low-redshift counterparts in terms of \Hb parameters, their \Feii emission is extremely low. Here we also aim at pinning down the possible causes of this \Feii weakness.

We describe the sample assembled for this work in Sec. \ref{sec:sample}. In Sec. \ref{sec:methods} and \ref{sec:results} we describe respectively the analyses performed on the sample and their outcomes. Lastly, the physical scenarios consistent with these observations are explored in Sec. \ref{sec:causes}. Our results are discussed in a broader context in \ref{sec:discussion}, while conclusions are drawn in \ref{sec:conclusions}. 

Throughout this work, we adopt a flat $\Lambda$CDM cosmology with $H_0 = 70$ km s$^{-1}$ Mpc$^{-1}$, $\Omega_{\Lambda}$ = 0.7, and $\Omega_{m}$ = 0.3.

\section{Sample}
\label{sec:sample}
The main goal of this work is to investigate the emission line properties in the rest-frame optical region including both the \Feii\footnote{Throughout this work when referring to \Feii we are referring to the \ion{Fe}{ii} blend in the range 4434--4684 \AA .} and the \Hb emission for high-$z$ AGN. Therefore, our sample of AGN was tailored by adopting the following criteria:
\begin{itemize}
    \item Previous identification as a broad line AGN. 
    \item Presence of a broad component in the \Hb profile.
    \item \textit{JWST} observations\footnote{As the \Hb line moves out of the K band at $z\gtrsim 4$ it becomes effectively inaccessible from the ground, thus our analysis in the high-redshift regime is mostly limited to objects observed with \textit{JWST}.} covering of the \Feii 4434--4684 \AA\, range.
\end{itemize}

For most of the sample we aimed at including high-$z$ sources ($z>5$), yet in three cases (namely JADES-028074, J-209777, XID-2028) we relaxed this criterion, as these sources offer the possibility to track the properties of low-luminosity AGN at intermediate redshifts. We list relevant information about the objects in the sample in Table \ref{tab:info}. 
In addition to the already known AGN with broad \Hb, we included three sources from the RUBIES survey (GO-4233; PI: A. de Graaff, \citealt{de2024rubies}) identified as broad line AGN. Lastly, with the aim of enriching the high-luminosity tail of this sample, we added the eight quasars from the ASPIRE survey (ID 2078, PI: F. Wang; \citealt{wang2023spectroscopic}) whose rest-frame optical properties have recently been analysed in \citet{yang2023spectroscopic}.

\begin{figure}
	\includegraphics[width=\columnwidth]{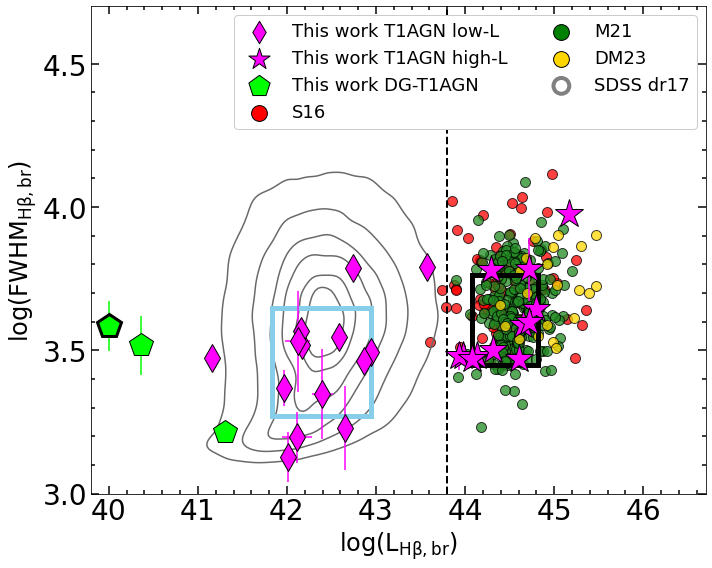}
    \caption{The $\log(\rm FWHM_{H\beta,br})-\log(\rm L_{H\beta,br})$ parameter space for our broad line AGN (T1AGN) and the reference samples (SDSS dr17, \citealt{shen2016rest} S16, \citealt{matthews2021placing} M21, \citealt{deconto2023high} DM23), where the SDSS dr17 AGN are shown with contours. We are also including the three broad line AGN in metal-poor dwarf galaxies (DG-T1AGN). The dashed line marks the threshold luminosity dividing the high- and the low-luminosity sub-samples. The actual $\log(\rm L_{H\beta,br})$ of SBS\_0335-052E (black thick edge) is 38.3, but it was shifted to 40.0 for the sake of a tighter image layout. The azure and black rectangles mark the regions adopted to define the control samples (Sec.\ref{sec:results}).}
    \label{fig:sel}
\end{figure}

To sum up, our sample is made of 26 high-redshift sources ($\sim$ 90\% of the sample is at $z>6$) out of which 14 have \Hb luminosity $\rm \log(L_{H\beta,br}/ (\ergs)) <$ 43.8. For the sake of simplicity, we will refer to these sources as the low-luminosity (or faint) AGN. The remaining 12 have instead $\rm \log(L_{\rm \Hb,br}/(\ergs)) >\,$43.8 and we will refer to them as the high-luminosity subsample (or quasars). Although the $\log(L_{\rm H\beta,br}/ \rm (erg \, s^{-1}) =\,$43.8 threshold is quite arbitrary, the two samples are fairly well separated in terms of \Hb luminosity, as shown in Fig.\ref{fig:sel}. The only object somewhat in-between the high- and the low-luminosity samples is XID-2028, having $\log(\rm L_{H\beta,br}/\rm (erg \, s^{-1}) =\,$43.6. Yet, its inclusion in either of the two samples does not significantly alter the average properties later discussed in the paper.

As a complement to our sample, we also introduce some reference samples at lower $z$ that were purposefully chosen in order to compare the properties of our objects. In particular, for what concerns the low-luminosity sub-sample, we chose the sources from the latest Sloan Digital Sky Survey (SDSS; \citealt{york2000sloan}) quasar catalogue, whose properties are described in \cite{wu2022catalog}. We select sources below $z$=0.8 with reliable \Hb measurements, by applying the quality cuts suggested in \citet{wu2022catalog} (see their Sec. 4 for details). 

For what regards the high-luminosity regime, a complication for finding suitable reference samples is given by the prevalence of quasars at redshift $z\sim 2-3$, the so-called "quasar epoch". At these redshifts, the \Hb region falls in the $H$ band which is not covered by large optical surveys. For this reason, we opted for objects targeted by near-infrared surveys. In particular, we adopted the samples described respectively in \citet{shen2016rest}, \cite{matthews2023gemini} and  \cite{deconto2023high}\footnote{Since it was not reported in their work, we derived $L_{H\beta,br}$ for the sources in \citet{deconto2023high} using the \ion{H}{$\beta$} equivalent width and the 5100 \AA\, luminosity, assuming a standard ratio between the luminosity of the continuum at the \ion{H}{$\beta$} location and at 5100 \AA\, of 1.05 (e.g. \citealt{vandenberk2001}).}. 
The \citet{shen2016rest} sample comprises 74 luminous quasars ($\rm L_{bol} = 10^{46.2-48.2} erg \, s^{-1}$) between $1.5 < z < 3.5$, observed with near-infrared (JHK) slit spectroscopy covering the rest-frame \Hb, \Feii and [\ion{O}{iii}] region.
The catalogue described in \citet{matthews2021placing} constitutes the Gemini Near Infrared Spectrograph-Distant Quasar Survey (GNIRS), containing a sample of 226 quasars between $1.5<z<3.5$ with infrared data covering the rest-frame optical/UV range. Lastly, the \citet{deconto2023high} sources are a similar sample of luminous ($\log(\rm L_{bol}/(\ergs))\sim$47.0-48.5) objects at $2.3<z<3.8$, with spectral coverage in the rest-frame optical band, purposefully observed with the goal of describing the \Feii and \Hb properties at intermediate $z$.

%%%%%%%%%%%%%%%%%%%%%%%%%%%%%%%%%%%%%%%%%%%%%%%%%%%%%%%%%%%%%%%%%%%%%%% INFO SAMPLE
\begin{table*}
\setlength{\tabcolsep}{4pt}
	\centering
	\begin{tabular}{ccccccc} % four columns, alignment for each
		\hline
		ID & RA & DEC & $z$ & Instrument & Configuration & ref. \\
        (1) & (2) & (3) & (4) & (5) & (6) & (7) \\
		\hline
		GS\_3073              & 57.078  & -27.884 & 5.555 &  NIRSpec/IFS &  G395M/F290LP - 0.25" & \citet{ubler2023ga} \\
		COS-ZS7               & 150.099 & 2.3436  & 7.145 &  NIRSpec/IFS &  G395M/F290LP - 0.50" & \citet{ubler2024ga} \\
	  	CEERS-01019           & 215.035 & 52.890  & 8.681 &  NIRSpec/MSA &  G395M/F290LP   & \citet{larson2023ceers} \\
	    JADES-000954          & 189.152 & 62.260  & 6.762 &  NIRSpec/MSA &  G395M/F290LP   & \citet{maiolino2023jades} \\
     	JADES-028074          & 189.065 & 62.234  & 2.261 &  NIRSpec/MSA &  G395M/F290LP   & \citet{juodvzbalis2024jades}\\
     	JADES-209777          &  53.156 & 27.776  & 3.711 &  NIRSpec/MSA &  235M/F170LP    & Juod{\v{z}}balis et al. in prep\\
        RUBIES-CEERS-49140    & 214.892 & 52.878  & 6.687 &  NIRSpec/MSA &  G395M/F290LP   & \citet{kocevski2024rise} \\
        RUBIES-CEERS-55604    & 214.982 & 52.956  & 6.985 &  NIRSpec/MSA &  G395M/F290LP   & \citet{kocevski2024rise} \\
	    UNCOVER-20466         &   3.640 & -30.386 & 8.502 &  NIRSpec/MSA &  PRISM/CLEAR    & \citet{kokorev2023uncover} \\
	    Abell2744-QSO1        &   3.604 & -30.382 & 7.045 &  NIRSpec/MSA &  PRISM/CLEAR    & \citet{furtak2024high} \\
	    XID-2028              & 150.547 &  1.619  & 1.593 &  NIRSpec/IFS &  G140H/F100LP - 0.25"  & \citet{cresci2023bubbles} \\
	    DELS J0411–0907       &  62.869 &  -9.130 & 6.820 &  NIRSpec/IFS &  G395M/F290LP - 0.35"-0.45"  & \citet{marshall2023ga} \\
	    VDES J0020–3653       &   5.131 & -36.895 & 6.855 &  NIRSpec/IFS &  G395M/F290LP - 0.35"-0.45"  & \citet{marshall2023ga} \\
	    PJ308-21              & 308.042 & -21.234 & 6.234 &  NIRSpec/IFS &  G395M/F290LP - 0.30" & \citet{loiacono2024quasar} \\
        J0100+2802            &  15.054 &  28.041 & 6.327 &  NIRCam/WFSS &  F356W             &  \citet{eilers2023eiger}\\

        \hline
        J010953.13-304726.30  & 17.471  & -30.791 & 6.790   &  NIRCam/WFSS & F356W      &  \citet{yang2023spectroscopic} \\
        J021847.04+000715.20  & 34.696  & 0.121   & 6.770   &  NIRCam/WFSS & F356W      &  \citet{yang2023spectroscopic} \\
        J022426.54-471129.40  & 36.111  & -47.192 & 6.522   &  NIRCam/WFSS & F356W      &  \citet{yang2023spectroscopic} \\
        J022601.87+030259.28  & 36.508  & 3.050   & 6.541   &  NIRCam/WFSS & F356W      &  \citet{yang2023spectroscopic} \\
        J024401.02-500853.70  & 41.004  & -50.148 & 6.731   &  NIRCam/WFSS & F356W      &  \citet{yang2023spectroscopic} \\
        J030516.92-315056.00  & 46.320  & -30.791 & 6.614   &  NIRCam/WFSS & F356W      &  \citet{yang2023spectroscopic} \\
        J200241.59-301321.69  & 300.673 & -30.223 & 6.688   &  NIRCam/WFSS & F356W      &  \citet{yang2023spectroscopic} \\
        J223255.15+293032.04  & 338.230 & 29.509  & 6.666   &  NIRCam/WFSS & F356W      &  \citet{yang2023spectroscopic} \\

		\hline
        RUBIES-EGS-8488     & 215.035 & 52.891  & 6.68 &  NIRSpec/MSA &  G395M/F290LP  &  This work \\
        RUBIES-EGS-948917   & 214.893 & 52.857  & 6.73 &  NIRSpec/MSA &  G395M/F290LP  &  This work  \\
        RUBIES-CEERS-966623$^\ddag$ & 214.876 & 52.881  & 8.35 &  NIRSpec/MSA &  G395M/F290LP  &  This work \\
        
        \hline
        SBS\_0335-052E      &  54.434  & -5.044  & 0.014 & FORS1  & 600B, 600R & \citet{izotov2009sbs} \\
        J102530.29+140207.3 & 156.376  & 14.035  & 0.101 & SDSS   & -- & \citet{izotov2008active} \\
        J104755.92+073951.2 & 161.983  & 7.664   & 0.168 & SDSS   & -- & \citet{izotov2008active} \\

		\hline
	\end{tabular}
\caption{Relevant information for the objects in our sample. Identifier (1), coordinates (2,3) and redshift (4). Column 5 reports the instrument employed for these observations. Column 6 described the configuration adopted for each observation. In the case of NIRSpec multi-shutter array (MSA) we report the disperser/filter coupling. In the case of integral field spectroscopy (IFS) we also report the radius of the circular extraction region in arcseconds. In the case of Wide Field Slitless Spectroscopy (WFSS) we report the filter employed for the observations analysed here. Column 7 highlights the reference papers for the spectral data employed in this work for each source. We also report the instrument and configuration for the spectra of AGN in dwarf galaxies.
\newline
$^\ddag$: already reported in \citet{kocevski2024rise}, but without the broad component in \Hb.}
\label{tab:info}
\end{table*}
%%%%%%%%%%%%%%%%%%%%%%%%%%%%%%%%%%%%%%%%%%%%%%%%%%%%%%%%%%%%%%%%%%%%%%%%

Lastly, alongside the main sample of low-luminosity \textit{JWST} objects, we also consider three extremely metal poor dwarf galaxies hosting a Type 1 AGN, identified by the presence of broad H$\alpha$ emission lines. These are SBS\_0335-052E (\citealt{izotov1990unusually}), J102530.29+140207.3 and J104755.92+073951.2 (\citealt{izotov2008active}) whose properties are extensively described in Sec.\ref{sec:dg_agn}. We included these sources as tentative very local analogues of the faint \textit{JWST} AGN since, as we will show briefly, they share a substantial set of similarities, in terms of optical lines and X-ray properties, with the high redshift sources.

In order to put our sources in the broader context of the emission line and accretion properties of local AGN, in Fig. \ref{fig:sel} we show our sample, together with the control samples on the $\log(\rm FWHM_{ H\beta,br})$--$\log(\rm L_{H\beta,br})$\footnote{When dealing with these parameters, for the sake of a lighter notation, we will simply report $\log(\rm FWHM_{\rm H\beta,br})$ and $\log(\rm L_{H\beta,br})$, rather than to the more correct -yet lengthy- notation $\log(\rm FWHM_{H\beta,br}/(km \, s^{-1}))$ and $\log(\rm L_{H\beta,br}/(erg \, s^{-1}))$.} plane. The choice of this parameters space presents several advantages. First, since the parameters describing the accretion process, such as the black hole mass and the bolometric luminosity, can be derived from these quantities (see e.g. \citealt{vestergaard2006determining, dalla2020sloan}), we expect objects residing in the same region of this parameter space to also share similar accretion parameters. Therefore, it is straightforward to consistently define the control samples. Secondly, since the parameter space is defined on the basis of observed quantities, the position of our sources therein is not subject to systematic uncertainties affecting the calibrations.

%%%%%%%%%%%%%%%%%%%%%%%%%%%%%%%%%%%%%%%%%%%%%%%%%%%%%%%%%%%%%%%
% PARAMETER SPACE
%%%%%%%%%%%%%%%%%%%%%%%%%%%%%%%%%%%%%%%%%%%%%%%%%%%%%%%%%%%%%%%

\subsection{Metal-poor X-ray weak local analogues of faint \textit{JWST} AGN}
\label{sec:dg_agn}

For comparison with galaxies in the local Universe, we included three metal-poor dwarf galaxies hosting low-luminosity Type 1 AGN (DG-T1AGN). 
The identification of the AGN in these sources has been first confirmed by the presence of a broad (FHWM $>$ 1000 km s$^{-1}$) \ion{H}{$\alpha$} emission line in \citet{izotov2008active}. In addition, more recent observations, 15 yr after the first ones (\citealt{burke2021agn}), confirmed the presence of the broad components in J102530.29+140207.3 and J104755.92+073951.2, thus excluding the supernovae shock scenario (\citealt{baldassare2016multi}). Here we also added a secure detection of a broad component in the \Hb profile.

From an observational perspective, there are several similarities between the known properties of the faint \textit{JWST} AGN and DG-T1AGN. Both these classes of objects have remarkably low NLR heavy element abundance with the highest measured oxygen abundance values in DG-T1AGN spanning $12+\log \mathrm{(O/H)}$ 7.3 and 8.0 (\citealt{izotov1999helium, burke2021agn}), and similar values having been detected in low-luminosity \textit{JWST} AGN (e.g. \citealt{harikane2023jwst, ubler2023ga, kocevski2023hidden,maiolino2023jades}). Another noticeable feature of these objects is the X-ray weakness of the AGN harboured therein (\citealt{thuan2004chandra, burke2021agn}). Additionally, there is also tentative evidence that the fraction of DG-T1AGN exhibiting absorption features in Balmer lines is higher than what is generally found in typical SDSS local AGN, just as reported in faint \textit{JWST} AGN (\citealt{matthee2024little, kocevski2024rise, juodvzbalis2024jades}). However, the current sample size is limited for solid conclusions. 
As we will show in the following Sections, and discuss more thoroughly in Sec.\ref{sec:discussion}, these properties, often observed also in the faint \textit{JWST} AGN, make AGN in metal poor dwarf galaxies an intriguing class of local analogues to compare the properties observed at remote cosmic distances.

\section{Methods}
\label{sec:methods}
In this section, we describe the techniques adopted to quantify the spectral properties of the sources in our sample.

\subsection{Spectral fits}
\label{sec:spectral_fits}
In order to access the spectral information embedded in the \textit{JWST} spectra, we performed a detailed spectroscopic analysis, focusing on a reliable determination of the \Feii and \Hb properties. To this end, we adopted a custom-made Python code, based on the IDL \textsc{MPFIT} package \citep{Markwardt2009}, which takes advantage of the Levenberg-Marquardt technique \citep{more1978levenberg} to solve the least-squares problem. The main emission lines were modelled by adopting different line profiles (Gaussian, Lorentzian). In the case of the broad \Hb line in quasars, we also use a broken power-law profile convolved with a Gaussian (e.g. \citealt{nagao2006}), as it proves more effective in describing the asymmetry often observed in the red side of the \Hb (see e.g. \citealt{deconto2023high}). The kinematics of [\ion{O}{iii}] has been tied to that of the narrow \Hb component, and the flux ratio between the 5007 \AA\, and the 4959 \AA\, components was fixed to three to one (\citet{osterbrock2006astrophysics}).
With the aim of reproducing the diversity of the \Feii emission, we included several spectral templates produced within the \textsc{Cloudy} environment (\citealt{ferland2013}). Specifically, we ran models on a grid spanning a wide range of physical parameters in order to broadly cover the parameter space of the \Feii emission. In particular, the models span the cloud Hydrogen density ($n_H$) range between $10^8 \leq n_e \leq 10^{14}$ cm$^{-3}$, photon ionising flux ($\phi$) between $10^{17} \leq \phi \leq 10^{23}$ cm$^{-2}$ s$^{-1}$, microturbulence velocity values $v_{turb}$ fixed to 0 and 100 km s$^{-1}$ and solar metal abundances. The continuum adopted is the default AGN continuum from \citet{mathews1987heats}. These models are then convolved with a Gaussian profile (the same for all the models as they are thought to represent co-spatial emission), shifted, and weighted during the fitting process. The parameters ultimately fitted for the \Feii templates are then the velocity dispersion and the shift of the Gaussian kernel as well as the weights of each template. 

Since we adopted a non-linear least square approach to perform the fit, the weights of the \Feii templates are constrained to be $>0$. This implies that, in the case of low signal and/or low \Feii emission, positive spikes of noise could be interpreted as actual iron emission, leading to an overestimate of the actual \Feii flux. In order to mitigate this issue we also adopted a non-parametric approach to estimate the \Feii contribution in the 4434--4684 \AA\, bump. In brief, we performed the spectral fit without the \Feii\ templates and focused on faithful modelling of the emission lines included in or close to this wavelength range (chiefly \ion{He}{i}$\lambda 4471$ and \ion{He}{ii}$\lambda 4686$). We then subtracted the best-fit models of these lines from the observed spectrum and considered the remaining emission to be ascribable to \Feii. If the integrated flux estimated by this second method was $<0$, we marked the measurement obtained via the \Feii templates (the parametric approach) as an upper limit.

We estimated the uncertainty on the best-fit parameters by adopting a Monte Carlo approach. Specifically, we performed the fit for 100 mock spectra for each source: the flux in every spectral channel was simulated by randomising the measured flux adopting a Gaussian noise whose amplitude was set by the uncertainty value in that spectral channel. After fitting every mock sample, we computed the distribution of the best-fit parameters and set the uncertainty as the standard deviation of the distribution, after applying a 3$\sigma$ clipping.

Examples of the spectral fits of both a low- and a low-luminosity object are shown in Fig.\ref{fig:fit_ex}. A complete gallery of all the fits is presented in Appendix \ref{app:fits_atlas}.

%%%%%%%%%%%%%%%%%%%%%%%%%%%%%%%%%%%%%%%%%%%%%%%%%%%%%%%%%%%%%%
% EX FITS

\begin{figure*}
  \centering
  \begin{minipage}[b]{0.46\textwidth}
    \includegraphics[width=\textwidth]{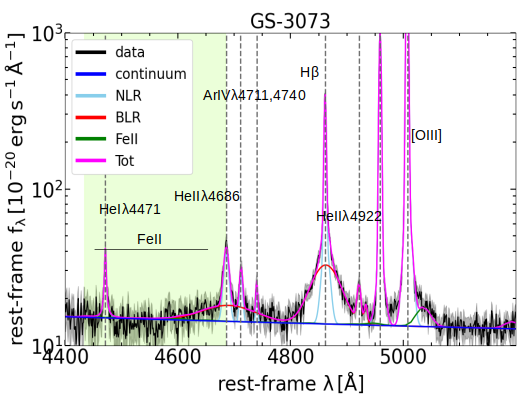}
  \end{minipage}
  \hfill
  \begin{minipage}[b]{0.47\textwidth}
    \includegraphics[width=\textwidth]{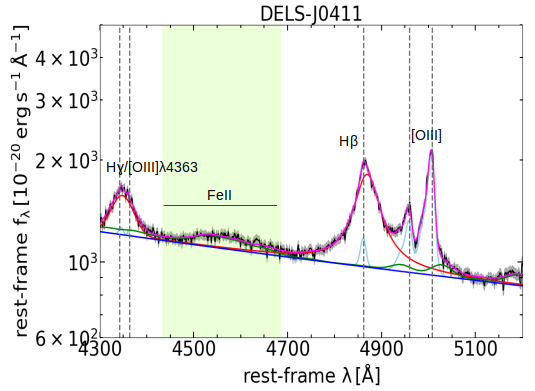}
  \end{minipage}
  \caption{Examples of the spectral fits of a low-luminosity (left) and a high-luminosity (right) AGN of the sample. The different components are colour-coded as stated in the legend. The uncertainty on the data is shown as a shaded area. The \Feii pseudo-continuum is highlighted in green. The vertical grey dashed lines mark the most prominent emission lines. All the spectral fits are shown in \ref{fig:fits_atlas}.}
  \label{fig:fit_ex}
\end{figure*}

%%%%%%%%%%%%%%%%%%%%%%%%%%%%%%%%%%%%%%%%%%%%%%%%%%%%%%%%%%%%%%

We report the spectral quantities of interest for the broad components, namely $\rm FWHM_{H\beta}$, \Rfe and $\rm \log L_{H\beta}$ \footnote{We note that the $\rm log L_{H\beta}$ are not corrected for intrinsic reddening within the BLR, yet this also applies to the control samples.} for each object in our sample in Table \ref{tbl:spec_pars}.

\setlength{\tabcolsep}{3pt}
\begin{table*}
	\centering
	\begin{tabular}{ccccccccc} % four columns, alignment for each
		\hline
		ID & FWHM $\rm H\beta_{br}$ & $\rm \log (L_{H\beta,br}$) & \Rfe & $\rm \log(L_{5100 \AA})$ & [OIII]$\rm /{H\beta_{na}}$ & O32 & EW[\ion{O}{iii}]  & 12+$\log$(O/H) \\ 
  
        & [$\rm km \, s^{-1}$] & $[\rm erg \, s^{-1}]$ &  & $[\rm  erg \, s^{-1}]$ 	&    &     & [\AA]  &          \\
        \\
		\hline
		GS\_3073              & 3525 $\pm$ 339  & 42.58 $\pm$ 0.04 & 0.03 $\pm$ 0.03 & 44.55 & 6.6 $\pm$ 0.2 & 12.8 $\pm$ 0.4 & 1084 $\pm$ 268   & 8.0$^{*}$ \\
		COS-ZS7               & 3289 $\pm$ 299  & 42.17 $\pm$ 0.04 & 0.09 $\pm$ 0.02 & 44.30 & 8.3 $\pm$ 0.2 & 14.4 $\pm$ 0.6 & 597  $\pm$ 11    & 8.0 $\pm$ 0.2 \\
	  	CEERS-01019           & 1571 $\pm$ 311  & 42.11 $\pm$ 0.17 & 0.22 $\pm$ 0.35 & 43.09 & 9.6 $\pm$ 1.4 & -              & 3251 $\pm$ 2499  & 8.1 $\pm$ 0.2 \\
	    JADES-000954          & 2338 $\pm$ 341  & 41.96 $\pm$ 0.06 & 0.19 $\pm$ 0.22 & 43.79 & 7.3 $\pm$ 0.8 & 13.3 $\pm$ 3.0 & 389  $\pm$ 23    & - \\
     	JADES-028074          & 3685 $\pm$ 136  & 42.16 $\pm$ 0.01 & 0.10 $\pm$ 0.01 & 43.45 & 5.0 $\pm$ 0.1 & 15.1 $\pm$ 0.4 & 254  $\pm$ 3     & 7.4 $\pm$ 0.2 \\
     	JADES-20977           & 6145 $\pm$ 692  & 42.74 $\pm$ 0.03 & 1.10 $\pm$ 0.05 & 44.75 & 1.8 $\pm$ 0.3 & -              & 20   $\pm$ 1     & - \\
        RUBIES-EGS-8488       & 1695 $\pm$ 506  & 42.65 $\pm$ 0.07 & 0.28 $\pm$ 0.04 & 43.82 & 9.1 $\pm$ 0.6         & 19.0 $\pm$ 5.0 & 1692 $\pm$ 165   & - \\
        RUBIES-CEERS-49140    & 3108 $\pm$ 135  & 42.95 $\pm$ 0.01 & 0.08 $\pm$ 0.03 & 44.44 & 26.9$^{**}$ $\pm$ 22.3 & -      & 99   $\pm$ 4     & - \\
        RUBIES-CEERS-55604    & 2905 $\pm$ 187  & 42.86 $\pm$ 0.02 & 0.14 $\pm$ 0.03 & 44.41 & 10.2 $\pm$ 2.1        & -              & 146  $\pm$ 6     & - \\
        RUBIES-EGS-948917     & 1344 $\pm$ 257  & 42.01 $\pm$ 0.06 & 0.41 $\pm$ 0.17 & 43.71 & 7.0 $\pm$ 0.6         & -              & 791  $\pm$ 123   & - \\
        RUBIES-CEERS-966323   & 2226 $\pm$ 972  & 42.40 $\pm$ 0.12 & 0.24 $\pm$ 0.19 & 43.87 & 4.2 $\pm$ 0.5         & -              & 218  $\pm$ 31    & - \\
	    UNCOVER-20466         & 3395 $\pm$ 1175 & 42.12 $\pm$ 0.14 & 0.02 $\pm$ 0.09 & 43.62 & 4.3 $\pm$ 3.2         & -              & 426  $\pm$ 153   & - \\
	    Abell2744-QSO1        & 2982 $\pm$ 267  & 41.16 $\pm$ 0.05 & 0.01 $\pm$ 0.03 & 43.25 & 14.8 $\pm$ 1.1        & -              & 4    $\pm$ 2     & - \\
	    XID-2028              & 6175 $\pm$ 218  & 43.57 $\pm$ 0.01 & 0.44 $\pm$ 0.05 & 45.35 & 11.4 $\pm$ 0.9        & -              & 80   $\pm$ 1     & - \\
	    DELS J0411–0907       & 3887 $\pm$ 76   & 44.65 $\pm$ 0.15 & 0.01 $\pm$ 0.05 & 46.38 & -                     & -              & -               & - \\
	    VDES J0020–3653       & 6058 $\pm$ 1083 & 44.71 $\pm$ 0.41 & 0.09 $\pm$ 0.05 & 46.24 & -                     & -              & -               & - \\
	    PJ308-21              & 4402 $\pm$ 22   & 44.79 $\pm$ 0.00 & 0.33 $\pm$ 0.00 & 46.41 & -                     & -              & -               & - \\
        J0100+2802            & 9477 $\pm$ 240  & 45.17 $\pm$ 0.01 & 0.26 $\pm$ 0.03 & 47.27 & -                     & -              & -               & - \\
        J010953.13-304726.30  & 3033 $\pm$ 117  & 43.98 $\pm$ 0.01 & 0.93 $\pm$ 0.05 & 45.81 & -                     & -              & -               & - \\
        J021847.04+000715.20  & 3030 $\pm$ 117  & 44.14 $\pm$ 0.02 & 1.45 $\pm$ 0.11 & 45.83 & -                     & -              & -               & - \\
        J022426.54-471129.40  & 3936 $\pm$ 78   & 44.72 $\pm$ 0.04 & 0.45 $\pm$ 0.05 & 46.47 & -                     & -              & -               & - \\
        J022601.87+030259.28  & 3131 $\pm$ 129  & 44.32 $\pm$ 0.01 & 1.11 $\pm$ 0.04 & 46.24 & -                     & -              & -               & - \\
        J024401.02-500853.70  & 2969 $\pm$ 158  & 44.60 $\pm$ 0.08 & 0.30 $\pm$ 0.05 & 46.18 & -                     & -              & -               & - \\
        J030516.92-315056.00  & 3020 $\pm$ 314  & 43.93 $\pm$ 0.03 & 1.78 $\pm$ 0.12 & 45.84 & -                     & -              & -               & - \\
        J200241.59-301321.69  & 2951 $\pm$ 79   & 44.07 $\pm$ 0.01 & 2.34 $\pm$ 0.08 & 46.14 & -                     & -              & -               & - \\
        J223255.15+293032.04  & 6044 $\pm$ 201  & 44.29 $\pm$ 0.01 & 1.61 $\pm$ 0.08 & 45.63 & -                     & -              & -               & - \\
        
		\hline
        SBS\_0335-052E      & 3845 $\pm$ 804 & 38.31 $\pm$ 0.02 & 0.05 $\pm$ 0.09 & 41.13 & 2.8 $\pm$ 0.2 & 14.8 $\pm$ 1.0 & 1044 $\pm$ 66 & 7.3 $\pm$ 0.2 \\
        J102530.29+140207.3 & 3287 $\pm$ 559 & 40.37 $\pm$ 0.07 & 0.28 $\pm$ 0.07 & 42.79 & 4.8 $\pm$ 0.2 & 8.9  $\pm$ 0.7 & 268  $\pm$ 5  & 7.5 $\pm$ 0.2 \\
        J104755.92+073951.2 & 1638 $\pm$ 100 & 41.30 $\pm$ 0.02 & 0.05 $\pm$ 0.02 & 43.23 & 6.4 $\pm$ 0.1 & 8.4  $\pm$ 0.2 & 869  $\pm$ 11 & 8.2 $\pm$ 0.2 \\
        \hline
	\end{tabular}
\caption{The broad lines spectral properties estimated from the spectral fits of our sample. All the FWHM have been corrected for the instrumental resolution. The uncertainty on $\log L_{5100}$ is negligible with respect to the others involved.\\
$^{*}$ The [\ion{O}{iii}]$\lambda$4363 is only marginally covered by the R2700 data. Here we report the value quoted in \citet{ubler2023ga}, where [\ion{O}{iii}]$\lambda$4363 kinematics had been tied to other narrow lines.\\
$^{**}$ RUBIES-CEERS-49140 presents an absorption feature close to the location of the narrow \Hb, which hampers a reliable measurement of the narrow component flux. Therefore, we mark this value as unreliable.}
\label{tbl:spec_pars}
\end{table*}

\subsection{Spectral stacks}
\label{sec:stacks}
In order to provide an immediate term of comparison at lower redshifts for the \textit{JWST} samples, we built spectral templates from the comparison samples described in Sec. \ref{sec:sample}.

Since the SDSS low-redshift sample covers a region of the adopted parameter space much wider than that spanned by our low-luminosity sample, we tailored a sample analogue to the \textit{JWST} one in the $\rm FWHM_{H\beta}$-$\rm \log(L_{H\beta, br})$ space. To this end, we selected a sample of SDSS AGN in the region of the parameter space defined as $[\rm \langle \log (L_{H\beta, br})\rangle \pm \sigma_{L_{H\beta, br}}, \langle \log(\rm FWHM_{H\beta})\rangle \pm \sigma_{FWHM_{H\beta, br}}]$, with the quantities in brackets being the mean values and the respective standard deviations for the low-luminosity sub-sample. The same selection was applied in the high-luminosity regime, with the only difference that here we only combined in the average spectrum the sources from \citet{matthews2021placing} and \citet{deconto2023high}, as \citet{shen2016rest} already provided a spectral composite of their full sample.

We then proceeded to produce the composite spectra for the four samples by performing the following steps:
\begin{itemize}
    \item Each spectrum was corrected for Galactic absorption assuming the value for the colour excess $E(B-V)$ available in the \citet{wu2022catalog} catalogue according to the \citealt{schlafly2011measuring} extinction maps and a \citet{fitzpatrick1999correcting} extinction curve. Then, the de-reddened spectra were shifted to the rest frame.

    \item All the spectra in the same sample were resampled, by means of linear interpolation, onto a fixed wavelength grid. Successively all the spectra were scaled by their 5100 \AA\, monochromatic flux, in order not to bias the stack towards the most luminous objects in each subsample.

    \item The final composite spectrum was obtained by taking the median value of the flux distribution in each spectral channel. The uncertainty on the median value was evaluated as the standard deviation in each spectral channel divided by the square root of the number of sources contributing to that channel. 
    
\end{itemize}

\begin{figure*}
  \centering
  \begin{minipage}[b]{0.47\textwidth}
    \includegraphics[width=\textwidth]{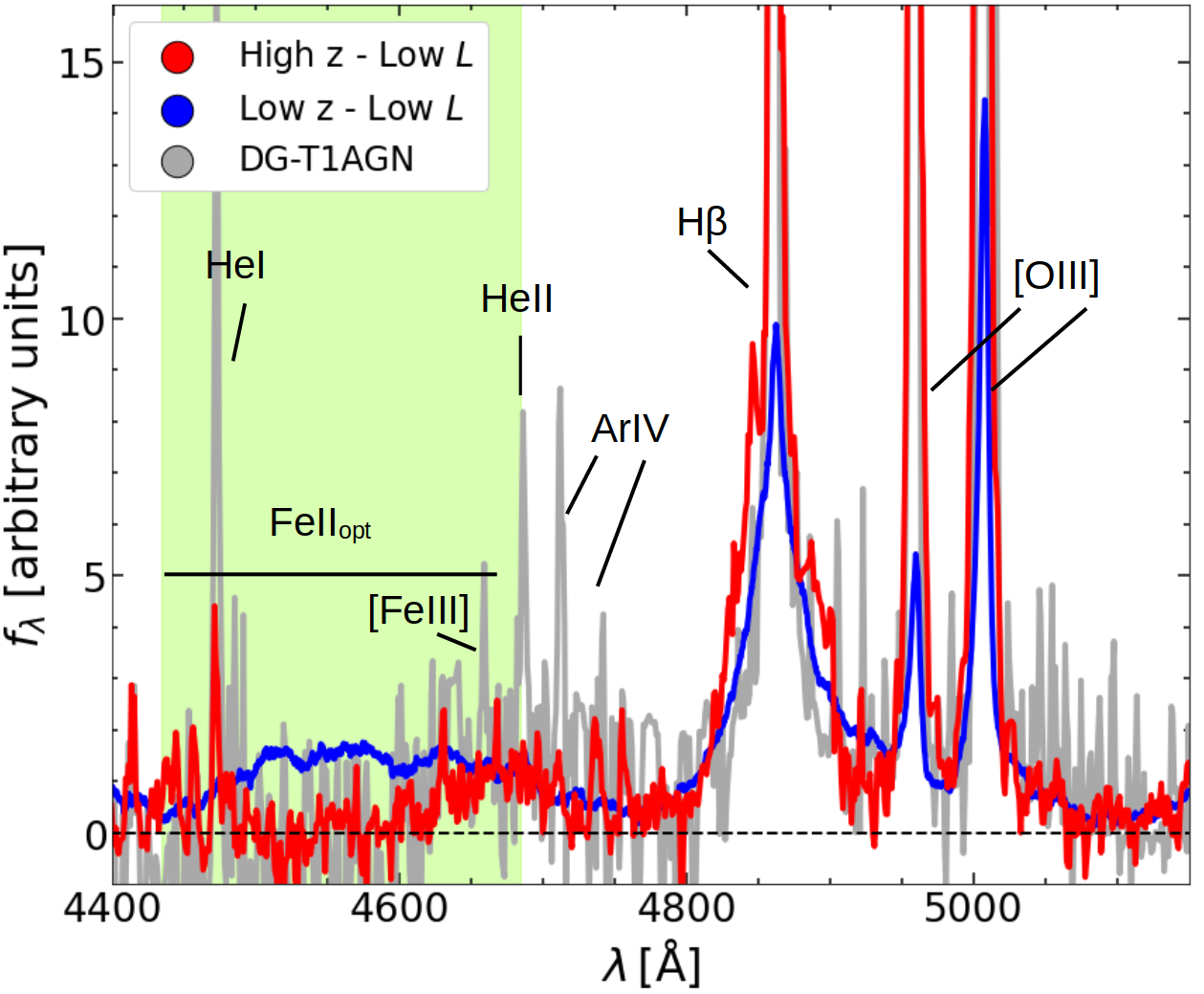}
  \end{minipage}
  \hfill
  \begin{minipage}[b]{0.47\textwidth}
    \includegraphics[width=\textwidth]{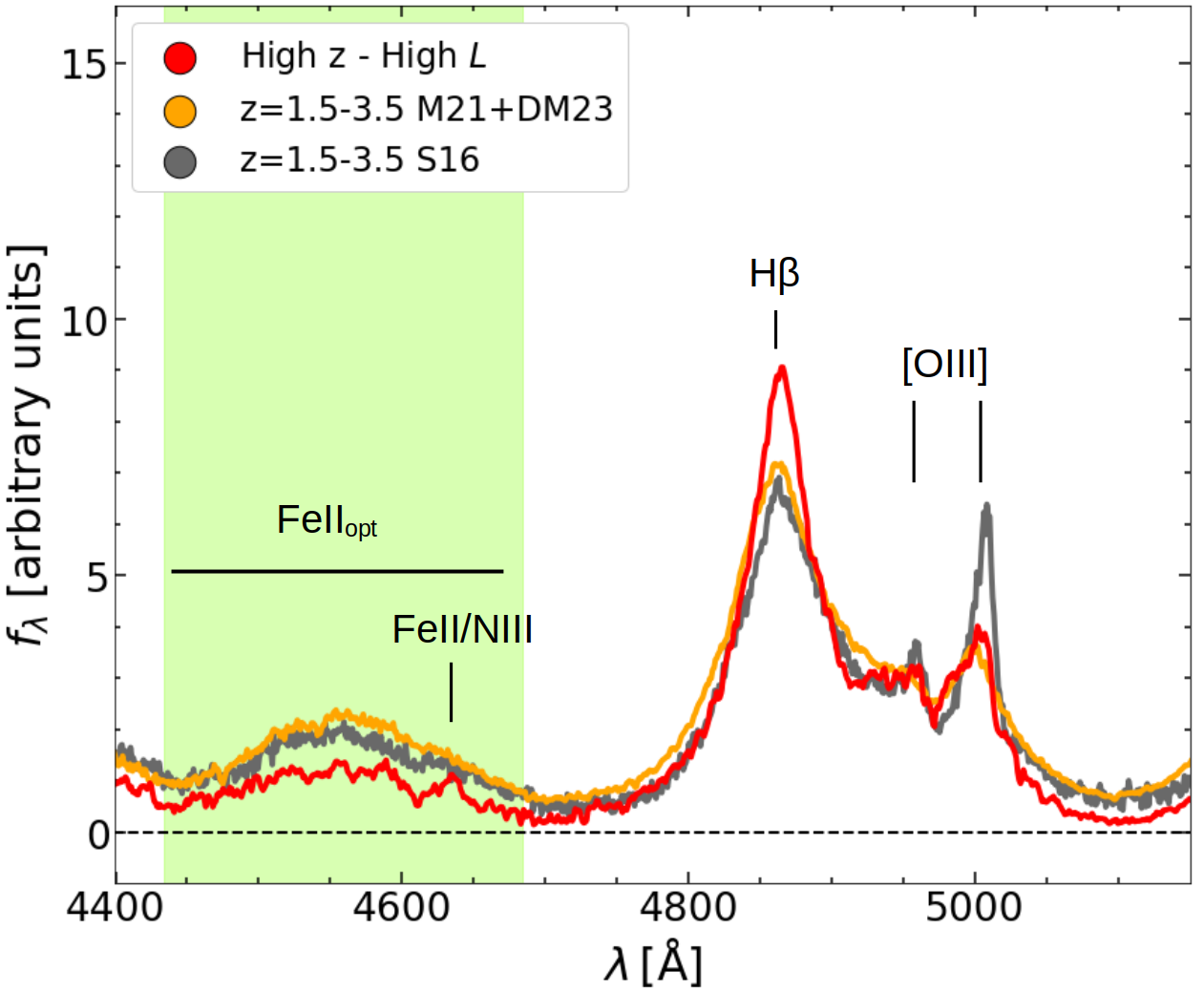}
  \end{minipage}
  \caption{Continuum-subtracted spectral composites of the low-luminosity (left) and high-luminosity (right) subsamples. The \Hb and [\ion{O}{iii}] narrow components have been cut for visualisation purposes. In the right panel we also show the composite quasar spectrum described in \citealt{shen2016rest}, made of 74 luminous quasars at $1.5<z<3.5$. The iron bump between 4434--4684 \AA\,(green shaded area) is evidently weaker in the \textit{JWST} low-luminosity sample than in the local AGN. However, the strength of the \Feii bump in the high-luminosity sample is comparable to the lower redshift analogue samples.}
  \label{fig:composites}
\end{figure*}

The same recipe was also adopted to produce the composite spectrum of the three DG-T1AGN.
The composite spectra produced as a result of this procedure, scaled by their broad \Hb fluxes and continuum-subtracted are shown in Fig.\ref{fig:composites}. 
There are several features clearly arising from this comparison. For what concerns the \Feii emission, the low-luminosity sample at high redshift exhibits a faint \Feii emission which translates into a \Rfe significantly lower than the local AGN counterparts. At the same time, the high-$z$ quasars present a slightly fainter \Feii than the other samples, yet their \Rfe is on average consistent with the other reference samples. The differences in terms of the \Rfe ratio are more quantitatively assessed in Section \ref{sec:results}. The lack of prominent \Feii in the low-luminosity objects can also be assessed by inspecting the fits of the composite spectra (Fig. \ref{fig:lowL_stack_o3_fit} and \ref{fig:composites_fits}), as well as the spectral fit of each source in Appendices \ref{app:lowL_spec_fits} and \ref{app:fits_atlas}.

% \textcolor{red}{Another noticeable feature is the stronger [\ion{O}{iii}] emission in the SDSS low-luminosity composite at low redshift. If we assume that the accretion disc powering these sources is not significantly different from that of the $z\sim 2-3$ quasars, as they reside in the same region of the parameter space, a possible explanation for this is an inclination effect. There is evidence for highly inclined sources to exhibit larger [\ion{O}{iii}] EW (\citealt{risaliti2011iii, bisogni2017inclination}). While in the local Universe it is easier to observe AGN with high inclination angles, the disc luminosity decreases by a factor cos $\theta$ (with $\theta$ being the inclination of the line of sight with respect to the axis of the disc), thus becoming increasingly more difficult to observe edge-on sources at higher $z$. Alternatively, since the small SDSS low-luminosity sample resides in a tail of the $L_{H\beta}$ distribution, it is also possible that these measurements are unreliable. Consequently, their high [\ion{O}{iii}] EW is simply consistent with those of the low-luminosity sample.}

For what regards the faint-AGN stack, we point out that the spectral fit reveals the presence of a significant\footnote{The significance of the inclusion of a broad component is assessed via the variation in the Bayesian Information Criterion (BIC). We consider as significant a decrease in the BIC by at least $\rm \mid \Delta BIC\mid > 10$} broad ($\sim$1000 $\rm km\,s^{-1}$, $\Delta$BIC = -67) component in the [\ion{O}{iii}] profile. This component is particularly interesting as it shows remarkable differences with respect to that observed in the SDSS control sample stacked spectrum. Firstly, this component is much fainter compared to the narrow component than that detected in the SDSS stack. The broad-to-narrow ratio in the SDSS composite is $\sim$0.45, while it is almost three times smaller in the \textit{JWST} one. Additionally, while the broad [\ion{O}{iii}] component is significantly blueshifted with respect to the core component in the SDSS stack, by $\sim$300 $\rm km\,s^{-1}$, the shift between the narrow and broad components in the faint \textit{JWST} AGN is consistent with zero within the uncertainties. Lastly, we also mention that the FWHM of the broad [\ion{O}{iii}] component is much smaller than the broad \Hb ($3731 \pm 133$ $\rm km \, s^{-1}$). Notably, this finding can be viewed as a clue against the interpretation of broad lines observed in these AGN as produced by star formation driven outflows (see e.g. \citealt{yue2024stacking, kokubo2024challenging}). As a matter of fact, we selected this sample of AGN only on the basis of the presence of broad lines, without any other kind of emission line diagnostic. If the mechanism producing the observed broad \Hb were star formation driven outflows, we would expect the \Hb and [\ion{O}{iii}] broad components to share similar kinematics. Although the high density might lead to suppression of [\ion{O}{iii}], causing the weakness of the broad component, the broad \Hb and the broad [\ion{O}{iii}] have significantly different kinematics (at 15$\sigma$). Therefore, this evidently points in the direction of a different origin for these two components.

% Another interesting difference between the SDSS and the high-redshift low-luminosity composite is the lack of the [\ion{O}{iii}] blue-wing tracing a ionised outflow in the local sample. As already pointed out in recent works (see e.g. \citealt{maiolino2023jades}), the population of high-redshift faint AGN seems to lack the broad blue-shifted wings routinely observed at lower redshift at all luminosities (see e.g. \citealt{shenho2014, bisogni2017inclination}). Interestingly, this feature is also missing in the DG-T1AGN composite spectrum. Yet, we notice that in this case the composite spectrum is made of only three objects.

%%%%%%%%%%%%%%%%%%%%%%%%%%%%%%%%%%%%%%%%%%%%%%%%%%%%%
% COMPOSITE COMPARISON

%%%%%%%%%%%%%%%%%%%%%%%%%%%%%%%%%%%%%%%%%%%%%%%%%%%%%%%%%%%%%%
\section{Results}
\label{sec:results}

\begin{figure}
	\includegraphics[width=\columnwidth]{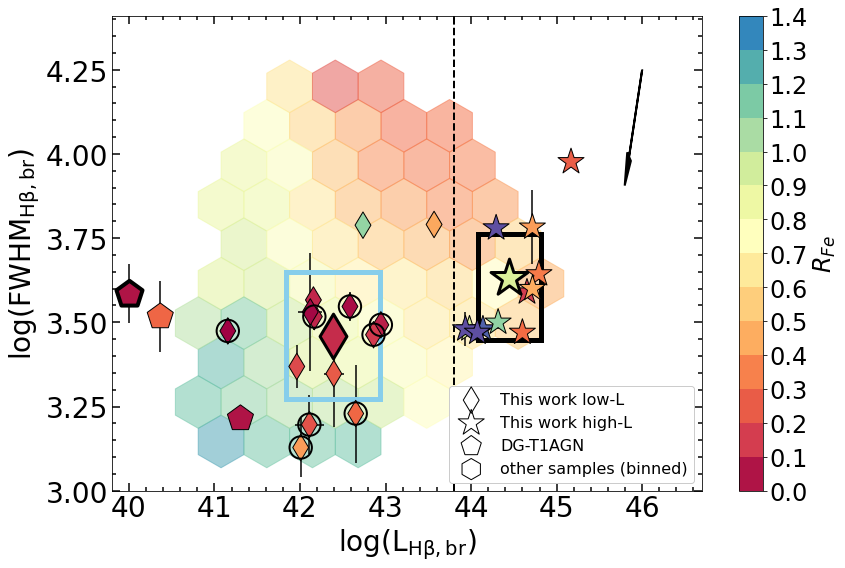}
    \caption{\Rfe-$\log(\rm FWHM_{H\beta})-\log(\rm L_{H\beta,br})$ diagram; specifically: $\log(\rm FWHM_{H\beta})$ versus $\log(\rm L_{H\beta,br})$ colour-coded by \Rfe. Objects marked with a circle represent \Rfe upper limits as defined in the text. The background 2D histogram is obtained by binning the control sample (with a minimum of 30 objects per bin); individual objects from \textit{JWST} are colour-coded with the same scale in \Rfe.
    Due to the large number of upper limits in the low-luminosity region, we also include the values derived from the composite spectra as symbols with thicker black edges. The coloured rectangles mark the regions where the control samples were drawn. The black solid arrow marks the direction of the PCA, as described in Sec. \ref{sec:acc_pars}. SBS\_0335-052E (black thick edge pentagon) has been shifted for a tighter image layout.}
    \label{fig:parspace_rfe}
\end{figure}

With the aim of understanding  the \Feii properties observed in our sample, in Fig. \ref{fig:parspace_rfe} we show again the FWHM$_{H\beta, br}$ vs L$_{H\beta, br}$ parameter space, adding this time the \Rfe in colour code. For a clearer visualisation of the surface, we binned the control samples in cells containing at least 30 objects. When the \textit{JWST} subsamples are compared to their control samples we observe a clear dichotomy, already foreseen in Fig. \ref{fig:composites}. The low-luminosity sample displays \Rfe much weaker than that observed for the same regions in the parameter space for the local control sample; the only notable exceptions are JADES-209777 and XID-2028, whose peculiarities will be discussed in Sec. \ref{sec:causes}. At the same time, the luminous quasars subsample is consistent, albeit with some scatter, with the expectations from the control sample at high luminosity.

Quantitatively, we estimated the significance of the difference in the \Rfe ratios using different statistical tests to assess what is the probability that the \Rfe distributions of the \textit{JWST} and control samples actually come from the same parent distribution as a null-hypothesis. In particular, we took advantage of the Kolmogorov-Smirnov test (KS; \citealt{hodges1958significance}), the Welch's t-test (W; \citealt{welch1947generalization}) suited for cases of small samples with unequal variances, and the Mann-Whitney U Test (MW; \citealt{mann1947test}). The results of these tests are reported in Table \ref{tbl:stat_tests}.

When the \textit{JWST} samples are matched to their respective control samples in the parameter space, all the tests confirm the different behaviour already seen in Fig.\ref{fig:composites}. The difference in the \Rfe ratios for the low-luminosity sample is extremely significant, with p-values $\lesssim 10^{-6}$. On the other hand, the \Rfe measured in the \textit{JWST} quasars is fully consistent with those measured at cosmic noon. In order to further strengthen these results against the possible effect of dust reddening within the BLR, we performed the same tests on a control sample chosen in order to match the average values of the de-reddened $\rm L_{H\beta,br}$. Again, the difference proved extremely significant (p-value$<10^{-6}$).

To conclude, we highlight that in the low-luminosity regime, we conservatively included also the X-ray detected AGN JADES-20977 and XID-2028, and considered the upper limits as actual measurements estimated via the parametric fits. The exclusion of these values would make the difference even stronger.

%%%%%%%%%%%%%%%%%%%%%%%%%%%%%%%%%%%%%%%%%%%%%%%%%%%%%%%%%%%%
% STATISTICAL TESTS
\setlength{\tabcolsep}{2pt}
\begin{table}
	\centering
	\begin{tabular}{cccc} 
		\hline
		sub-samples & KS & W & MW \\ 
		\hline
		\textit{J} low-$L$ / \textit{J} high-$L$ (1) & $2\times 10^{-4}$ & $5\times 10^{-3}$  & $10^{-4}$ \\
		\textit{J} low-$L$ / CS low-$L$ (2) & $4\times 10^{-9}$ & $9\times 10^{-7}$ & $10^{-8}$ \\
        \textit{J} high-$L$ / CS high-$L$ $z=2-3$ (3) & 0.120 & 0.154 & 0.279 \\
        \hline
        \textit{J} low-$L$ / CS low-$L$ de-red (4) & $3\times 10^{-9}$ & $2\times 10^{-6}$ & $4\times 10^{-8}$ \\
        \hline
        \textit{J} low-$L$ / [\ion{O}{iii}] CS (5) & 0.035 & 0.050 & 0.011 \\
        \hline
        \hline
	\end{tabular}
\caption{p-values of the statistical tests performed on the \jwst low- and high-luminosity samples (\textit{J}) against their respective control samples (CS; 1,2,3). Rows 4 and 5 denote respectively the same tests performed on the samples matching the de-reddened $\rm L_{H\beta}$ and the [\ion{O}{iii}] control sample described in \ref{sec:o3_Rfe_4dev1}.}
\label{tbl:stat_tests}
\end{table}
%%%%%%%%%%%%%%%%%%%%%%%%%%%%%%%%%%%%%%%%%%%%%%%%%%%%%%%%%%%%

% Example figure

\begin{figure*}
  \centering
  \begin{minipage}[b]{0.49\textwidth}
    \includegraphics[width=\textwidth]{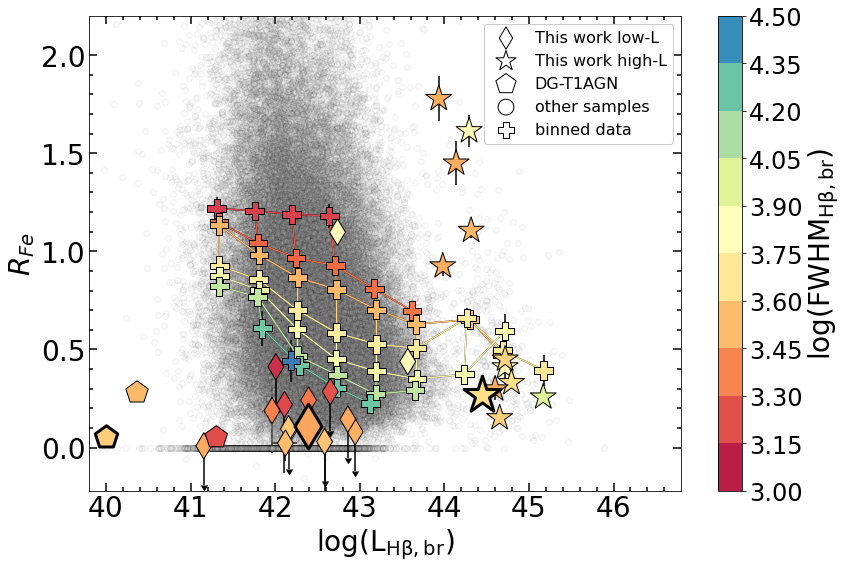}
  \end{minipage}
  \hfill
  \begin{minipage}[b]{0.49\textwidth}
    \includegraphics[width=\textwidth]{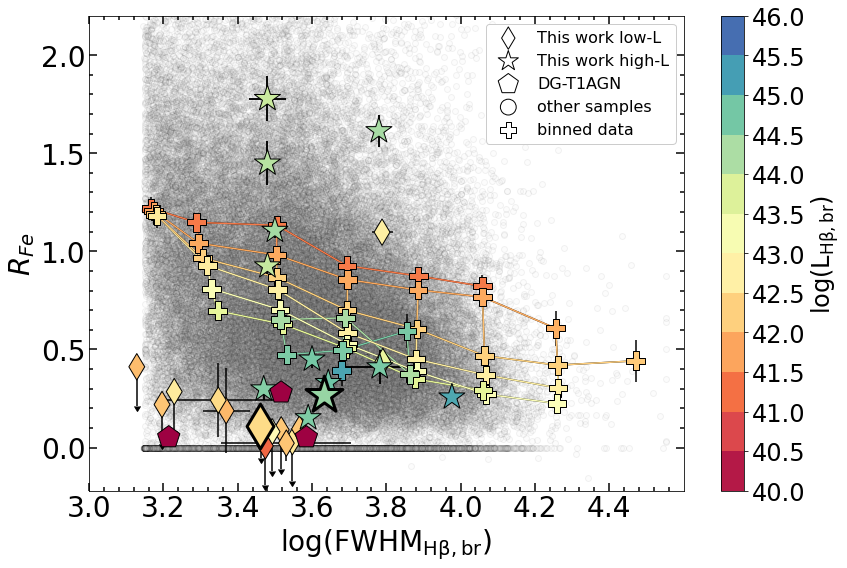}
  \end{minipage}
  \caption{Orthogonal sections of the $\log(\rm FWHM_{H\beta})-\log(L_{H\beta})-$\Rfe parameter space; specifically (left) \Rfe versus $\log(L_{H\beta})$ in (colour-coded) bins of $\rm FWHM_{H\beta}$, and (right) \Rfe versus $\rm FWHM_{H\beta}$ in (colour-coded) bins of $\log(L_{\rm H\beta})$. The thick-edged diamond and star represent the values derived from the composite spectra. The colour-coded lines represent the third quantity evaluated on the binned sources of the control samples. SBS\_0335-052E (black thick edge pentagon) has been shifted for a tighter image layout. It is clear that, while the high-luminosity objects of the sample reside -albeit with some scatter- in the expected locus of the parameter space according to the control sample, the low-luminosity objects exhibit far lower \Rfe. Interestingly, two low-luminosity objects (JADES-029777 and XID-2028) detach from the bulk of the faint sample, being close to the \Rfe expected. As we discuss further in Sec.\ref{sec:Xray_perspective}, these are the two only faint AGN which are X-ray detected.}
  \label{fig:parspace_sections}
\end{figure*}

As a more straightforward way to notice the region occupied by the \jwst sources with respect to the bulk of the lower $z$ samples, we also present two side views on the $\rm FWHM$-$\rm L_{H\beta}$-\Rfe parameter space in Fig. \ref{fig:parspace_sections}. Specifically, the left panel shows \Rfe versus $\rm L_{H\beta}$ in (colour-coded) bins of $\rm FWHM$, while the right panel shows \Rfe versus $\rm FWHM$ in (colour-coded) bins of $\rm L_{H\beta}$. It is clear that the low-luminosity sample exhibits systematically lower \Rfe than that expected for the corresponding region in the parameter space based on lower-$z$ data.

%%%%%%%%%%%%%%%%%%%%%%%%%%%%%%%%%%%%%%%%%%%%%%%%%%%%%
% PARAMETER SPACE SECTIONS

%%%%%%%%%%%%%%%%%%%%%%%%%%%%%%%%%%%%%%%%%%%%%%%%%%%%%%%%%%
%%%%%%%%%%%%%%%%%%%%%%%%%%%%%%%%%%%%%%%%%%%%%%%%%%%%%%%%%%%%

\section{Possible causes of the observed differences}
\label{sec:causes}

There is no consensus about the main drivers of the correlations involving the strength of the \Rfe ratio falling in the ensemble of the 4DE1. 

It is generally thought that the \Rfe-$\rm FWHM_{H\beta,br}$ anti-correlation results from the combined effect of the accretion rate and inclination of the line of sight with respect to the axis of the accretion disc powering the AGN. Yet, it is unlikely that the $\lambda_{Edd}$ is \textit{per se} the main driver of the 4DE1 correlation. The $\lambda_{Edd}$ increases perpendicularly to the main sequence, and objects with low Eddington ratios produce very different \Rfe ratios. The same spread in terms of \Rfe is observed in samples allegedly made of sources accreting at high-Eddington ratios such as the SEAMBH sample (see e.g. \citealt{du2018supermassive}). Also, from a theoretical perspective, the inclusion of a physically motivated warm X-ray corona has been observed to loosen the dependence of the \Rfe on the $\lambda_{Edd}$ (\citealt{panda2019cloudy}). 

Lastly, we must bear in mind that both the $\rm FWHM_{H\beta}$ and the Eddington ratio are easy to estimate parameters, useful when it comes to roughly describing the accretion properties of our sample, but they are not the physical quantities ultimately governing the micro-physics of \Feii and \Hb emissivity. Their effect on the \Rfe ratio comes more subtly in the form, for instance, of a dependence of the \Rfe on the SED, which, in turn depends on the accretion properties of the AGN. 

In this section, we present a series of tests aiming at exploring the possible drivers of the observed properties, in both our sample and the reference samples.

\subsection{Are we probing an extreme tail of the 4DE1?}
\label{sec:o3_Rfe_4dev1}
Within the 4DE1 set of correlations, it is well known that objects with stronger \Rfe exhibit weaker [\ion{O}{iii}] emission lines and vice-versa (\citealt{boroson1992emission}). Since our faint AGN show quite strong [\ion{O}{iii}] emission ($\sim \langle 600\rangle$ \AA), it is therefore legitimate to question whether the \Rfe weakness detected in these sources might be interpreted as the high-EW[\ion{O}{iii}] and low-\Rfe end of the eigenvector 1. We explored this possibility and presented several arguments against this interpretation.

Firstly, we note that the faint \textit{JWST} AGN are not consistent with the expectations for local (i.e. SDSS) sources in the same region of the $\log$(EW[\ion{O}{iii}])--\Rfe plane. This can be seen in Fig.\ref{fig:O3_Rfe}. There we highlight in pink the local sources -this time not selected from the $\log(\rm FWHM_{H\beta})-\log(\rm L_{H\beta,br})$ control sample- within 0.2 dex from the mean value of the \textit{JWST} low-luminosity sources. We tested whether the \Rfe estimated from the \textit{JWST} and the SDSS EW[\ion{O}{iii}]-matched sources could be compatible with coming from the same parent distribution, adopting the same statistical tests described in Sec. \ref{sec:results}. The \Rfe distribution for the \textit{JWST} sources resulted significantly different (at $\gtrsim$95\% confidence level, depending on the test, see Table \ref{tbl:stat_tests}) from the SDSS one. Here we also note that we made the very conservative assumption of considering the upper limits as actual measurements\footnote{Also the choice of the 0.2 dex interval in EW[\ion{O}{iii}] is conservative. The combination of lower \Rfe values with increasing EW[\ion{O}{iii}], combined with high EW[\ion{O}{iii}] sources becoming increasingly rarer, would produce an higher mean value of $\log(R_{Fe})$ for the control sample and therefore a stronger discrepancy if we considered sources between $[\rm \langle \log (EW[OIII])\rangle \pm \sigma_{EW[OIII]}]$.}. In Fig.\ref{fig:O3_Rfe} we also show that the average \Rfe values for the faint \textit{JWST} sources are far below ($>2.5 \sigma$) the best-fit relation for the SDSS full sample. There we estimated the average \Rfe in multiple ways: including the upper limits as actual measurements (maroon), extracting the non-detected values from a uniform distribution between the SDSS lowest 1$^{st}$ percentile and the upper limits (red), and using the value derived from the spectral fit of the stack (magenta).

Furthermore, other arguments can be brought against the faint \textit{JWST} sources fitting within the 4DE1 framework, which are the average Eddington ratio and the inclination.
The Eddington ratio, albeit the already mentioned caveats, is generally observed to correlate, with \Rfe (e.g. \citealt{shenho2014}). These sources exhibit, on average, the same Eddington ratio as their low-$z$ counterparts, having $\langle \lambda_{\rm Edd}\rangle \sim$ 0.14 ($\langle \lambda_{\rm Edd}\rangle_{\rm SDSS} \sim$ 0.11) adopting the \citet{vestergaard2006determining} calibration for the black hole mass and the \citet{shen2011} for the bolometric luminosity. Although the validity of the local calibrations in these so different environments is questionable, the \Rfe is far lower than the expectations.
The other mechanism classically invoked to explain the 4DE1 trends is the inclination of the line of sight. Objects with large [\ion{O}{iii}] are thought to be observed under large viewing angles (e.g. \citealt{risaliti2011iii, bisogni2017inclination}), while a more face-on inclination decreases the [\ion{O}{iii}] EW while increasing \Rfe. However, the large fraction of \textit{JWST} low-luminosity objects with large (EW[\ion{O}{iii}]$\gtrsim 100$ \AA) does not seem reconcilable with the local one ($\sim 4\%$ in Fig. \ref{fig:O3_Rfe}). At the same time, the inclination hypothesis to explain the large [\ion{O}{iii}] would leave room for the other -perhaps even more striking- question: where are all the low-inclination, low EW[\ion{O}{iii}] sources at these redshifts?

\begin{figure}
	\includegraphics[width=\columnwidth]{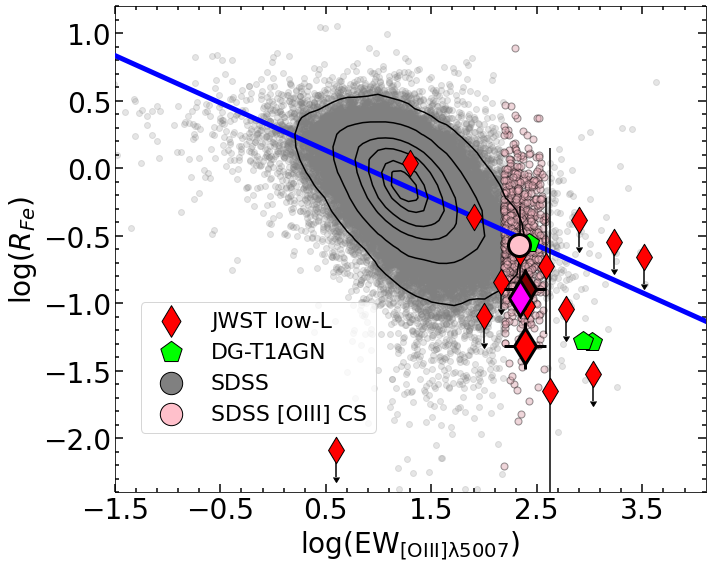}
    \caption{The $\log$(EW[\ion{O}{iii}])--$\log$(\Rfe) section of the 4DE1. The pink dots mark the SDSS objects selected as a control sample. The blue line highlights the best linear fit to the SDSS data. The thick-edged diamonds represent the average values for the faint \textit{JWST} AGN as described in the text.}
    \label{fig:ewo3_Rfe}
\end{figure}

Lastly, we mention that all the considerations made so far subsumed the AGN nature of the [\ion{O}{iii}] emission. This could be not the case, as for these sources a significant fraction of the [\ion{O}{iii}] emission could be ascribable to star formation. Indeed, star-forming galaxies and the AGN discovered at these redshifts overlap in the classical BPT diagrams (see e.g. \citealt{maiolino2023jades, kocevski2023hidden, harikane2023jwst}). Therefore, the EWs computed here would only be an upper limit to those actually coming from the AGN. Disentangling the AGN contribution from the star-formation would result in a shift leftwards of the average [\ion{O}{iii}] values in Fig. \ref{fig:ewo3_Rfe}, thus making the inconsistency with the SDSS 4DE1 even stronger.

%%%%%%%%%%%%%%%%%%%%%%%%%%%%%%%%%%%%%%%%%%%%%%%%%%%%%%%%%%%%%%%
% LOG EW[OIII] - LOG RFe SPACE
%%%%%%%%%%%%%%%%%%%%%%%%%%%%%%%%%%%%%%%%%%%%%%%%%%%%%%%%%%%%%%%

\subsection{Accretion parameters}
\label{sec:acc_pars}
The black hole mass and the accretion rate (or its closely related observable, the luminosity) are thought to ultimately rule, with a few other parameters, the shape and the emissivity of the accretion disc in AGN. These quantities are expected to be, therefore, the strongest drivers in the changes of the AGN SED, mainly responsible for the photoionisation in these sources. Hence, an obvious point is to assess whether the observed difference in \Rfe could be ascribed to changes in one or both of these parameters.

Nonetheless, there are some complications: the first is the fact that the evaluation of the $M_{BH}$ comes with a significant systematic uncertainty due to single-epoch calibrations (between 0.3-0.5 dex, see e.g. \citealt{shen2013mass, dalla2020sloan}). In this work, we adhered to the \citealt{vestergaard2006determining} prescription, also employed in \citet{wu2022catalog} for a fair comparison with the control sample.

Moreover, there is not a well-defined relation between the \Rfe and any of the accretion parameters, but the $\lambda_{Edd}$, with which there is a correlation, although \citet{panda2019cloudy} argued that this is more of an observational rather than due to an intrinsic effect. In the effort to understand the evolution of the \Rfe within our parameter space, we took advantage of a partial correlation (PC) analysis. This technique allows us to estimate the correlation between two quantities (in our case the \Rfe and the main axes of the parameter space) while holding fixed any other components.
In particular, given the three quantities (A,B,C), the partial correlation coefficient between A and B at fixed C can be expressed as:

\begin{equation}
    \rho_{AB \mid C} = \frac{\rho_{AB} - \rho_{AC} \rho_{BC}}{\sqrt{1 - \rho_{AC}^2} \cdot \sqrt{1 - \rho_{BC}^2}}
\end{equation}
where $\rho_{AB \mid C}$ denotes the Spearman rank correlation index between A and B, while keeping C constant. We also note that a monotonic trend is a requirement for the PCA to provide meaningful results, and this is mostly true as the \Rfe tends to increase with decreasing $FWHM_{H\beta}$ and increasing $L_{H\beta,br}$.

We used the PCA values to define a gradient arrow in Fig.\ref{fig:parspace_rfe}, whose inclination is defined as:
\begin{equation}
    \rm \tan \theta = \frac{\rho_{R_{Fe} \, FWHM_{\rm H\beta,br} \mid L_{\rm H\beta,br}}} { \rho_{R_{Fe} \, L_{\rm H\beta, br} \mid FWHM_{\rm H\beta,br}}}
\end{equation}

The more the arrow is aligned with an axis, the stronger is the correlation with that quantity. The correlation with the $\rm FWHM_{H\beta}$ is significantly stronger than with the $\rm L_{H\beta,br}$, yet both of them yield low, yet significant, partial correlation coefficients ($\rm \rho_{R_{Fe} \, L_{\rm H\beta,br} \mid  FWHM_{\rm \Hb,br}}  = -0.166$ with p-value $<10^{-6}$ and $\rm \rho_{R_{Fe} \, FWHM_{\rm H\beta} \mid L_{\rm H\beta,br}} = -0.281$ with p-value $<10^{-6}$).
At fixed $\rm L_{\rm H\beta,br}$, the increase in $\rm FWHM_{\rm H\beta}$ can be readily interpreted as an increase in $M_{BH}$, as a consequence of the \ion{H}{$\beta$} virial relation (see e.g. \citealt{dalla2020sloan} and references therein). At the same time, $\rm L_{H\beta,br}$ is tightly related to the 5100 \AA\, luminosity (see e.g. \citealt{kaspi2005relationship, dalla2020sloan}) which in turn is a good proxy to the bolometric luminosity (\citealt{richards2006spectral}). Yet, none of these two parameters seems to be a strong driver of the observed changes in \Rfe. Additionally, we recall that the difference evaluated in Section \ref{sec:results} has been computed between the \textit{JWST} low-luminosity objects and their local counterparts in the parameter space, which is tightly related to the accretion parameters. Therefore, we do not expect either the black hole mass nor the luminosity to be driving the observed difference, under the non-trivial assumption that the local calibrations apply also for this class of AGN. The same consideration applies to the Eddington ratio as well.

\subsection{The effect of metallicity}

A trivial factor which could be responsible for the weakness of the \Feii bump is the metallicity. Here we explore what we can infer about the gas-phase metallicity for the low-luminosity sources by employing line diagnostic ratios.

Extensive work based on CLOUDY simulations showed that it is possible to reproduce the diversity in the optical \Rfe of the low-$z$ SDSS sample allowing for a super-solar metallicity [1 Z$_{\odot}$-10 Z$_{\odot}$] while keeping the other BLR parameters, such as the density and the column density, within typical values (\citealt{panda2018modeling}). However, there is evidence for the faint objects in this early stages of the Universe not to have reached the chemical maturity observed in more local sources, at least in the NLR (\citealt{curti2023chemical, isobe2023redshift, maiolino2023jades, ubler2023ga, ubler2024ga, kocevski2024rise}). This points in the direction of low metallicity as a possible driver of the low \Rfe ratios observed in our sample. 

\subsubsection{Preliminary considerations}
The most straightforward way to estimate the metallicity of the BLR gas in these sources would be to employ high ionisation line ratios such as \ion{N}{v}/\ion{C}{iv} or (\ion{Si}{iv}+\ion{O}{iv})/\ion{C}{iv} (see e.g. \citealt{lai2022chemical} and references therein). Unfortunately, this wavelength range is not accessible within our spectra. Alternatively, we could infer the metallicity from the readily available rest-optical lines in the NLR, but translating this measurement into a BLR metallicity would require a model of the chemical enrichment and of the interplay between these two spatially different regions at these early cosmic epochs.

From an observational point of view, it is well known that the BLR and NLR metallicities are linked, with the former reaching generally higher metallicity ($Z\gtrsim Z_{\odot}$, e.g. \citealt{wang2022metallicity}) at early cosmic epochs. However, this link between the NLR and BLR metallicities implies that even if we managed to get reliable estimates of the NLR metallicity for the low-luminosity objects, these would only set a \textit{lower limit} to the corresponding BLR metallicity.

In addition, estimating the metallicity of the NLR in AGN is not a well-established procedure (see e.g. \citealt{dors2020chemical_a} for a compilation of the few attempts made in this direction) and efforts in this field have been mainly directed towards Type 2 AGN (\citealt{thomas2019mass, dors2020chemical_b, li2024mass}). An example of complications in estimating the AGN metallicity is the higher fraction of $\rm O^{3+}$ (whose corresponding transitions are undetectable in the optical range). This is associated with the harder AGN continuum and the effect of temperature inhomogeneities in the NLRs (\citealt{riffel2021electron}).

For our sample, the most straightforward way to estimate the metallicity of the NLR of our AGN would be via the electron temperature method (\citealt{smith1975spectrophotometric}), which is based on the use of the [\ion{O}{iii}]$\lambda$4363 auroral line. Unfortunately, this line is detected above 2$\sigma$ in only 4 out of 14 low-luminosity AGN. In comparison, strong-line methods would usually require detection of low ionization lines such as [\ion{N}{ii}]$\lambda \lambda$6548,6584, which are outside of the wavelength coverage for most of the sample, or barely detectable as outshined by the broad \ion{H}{$\alpha$}. We also recall that it is quite challenging to detect the [\ion{O}{iii}]$\lambda$4363 line in local Type 1 AGN to compare to the \textit{JWST} results, as it is generally faint and not easy to deblend from the narrow and broad \ion{H}{$\gamma$} and the \Feii pseudo-continuum (\citealt{baskin2005controls}). This leads to the additional complication of having a comparison sample (the SDSS local AGN catalogue) where metallicity measurements would be biased towards high [\ion{O}{iii}]$\lambda$4363 emitters. In addition, it is well known that the metallicity of \ion{H}{ii} regions and star-forming galaxies, directly estimated via this method, are systematically lower than those produced adopting calibrations from photoionisation modelling by 0.1-0.4 dex (see e.g. \citealt{kennicutt2003composition, dors2005abundance, lopez2007localized, kewley2008metallicity, marconi2024homerun}). This difference is even exacerbated in the case of AGN with the oxygen abundance being underestimated on average by 0.6--0.8 dex (\citealt{dors2015central, dors2020chemical_a}).

Lastly, we note that the the oxygen abundance estimated via the $T_e$ method or other calibrations (e.g. \citealt{storchi1998chemical, castro2017new}) does not necessarily trace the actual iron abundance, unless a chemical enrichment model is assumed. Indeed, high-redshift objects have been observed to display gas-phase oxygen abundances consistent with those observed locally (e.g. \citealt{arellano2022first, jones2023early}), as a consequence of the enrichment by core-collapse supernovae, while the total Fe abundance is expected to be significantly lower, having this element a delayed enrichment contribution from Type Ia SNe. An additional complication is the fact that Fe is heavily depleted onto dust grains (see e.g. \citealt{jenkins2009unified, shields2010}). Although, within the standard evolutionary picture, the effect of depletion onto dust grains in the redshift range spanned by our sources is not expected to be a crucial channel to suppress the \Feii emission, as it might happen at lower redshift (\citealt{shields2010}), more detailed studies are definitely needed to assess this possibility.

\subsubsection{Oxygen abundance}
Notwithstanding all these considerations, for 4/14 objects we detected [\ion{O}{iii}]$\lambda$4363 with SNR$\geq$2. Assuming Gaussian noise, and the redshift of each source being well determined, the 1-tailed probability of a false 2-sigma detection of [\ion{O}{iii}]$\lambda$4363 is 0.022. Supposing, in the worst case scenario, that the [\ion{O}{iii}]$\lambda$4363 line were actually absent in all sources, the joint probability of having 4 or more detections only due to positive statistical fluctuations would be $\simeq 1\times 10^{-4}$. In the following, we assume all detections to be real, and proceed to estimate the gas-phase oxygen abundance.

We derived the electron temperature ($T_e$) using the \textsc{pyneb} \textit{getTemDen} routine (\citealt{luridiana2015pyneb}). In most of these objects, we could not access the typical optical density indicators such as the [\ion{O}{ii}]$\lambda \lambda$3726,3729 or the [\ion{S}{ii}]$\lambda \lambda$6716,6731 doublets. Actually, the spectral fit of the [\ion{O}{ii}]$\lambda \lambda$3726,3728 (unresolved) doublet was successfully carried out only in five objects, and in two cases (COS-ZS7 and JADES-028074) this happened with simultaneous detection of the [\ion{O}{iii}]$\lambda$4363 line. To compensate for this, we computed the electron temperature using an equispaced grid of densities from 10$^1$ cm$^{-3}$ to 10$^5$ cm$^{-3}$. For each density value we evaluated the corresponding temperature of the high ionisation region ($t_3$) using the ratio between the [\ion{O}{iii}]$\lambda$5007 and the [\ion{O}{iii}]$\lambda$4363 emission lines (the O33 ratio) using the \textit{getTemDen} routine\footnote{Line fluxes were corrected for the reddening (if present), evaluated from the Balmer ratio, assuming the SMC extinction curve from \citet{gordon2016panchromatic}.}. Finally, we computed the oxygen abundances employing the \textsc{pyneb} \textit{getIonAbundance} routine. The uncertainty associated with the abundance measurements was derived as the $\rm 16^{th}-84^{th}$ semi-interpercentile range of the abundance distribution obtained varying the density. For the objects without [\ion{O}{ii}]$\lambda$3728 detection we assumed the average [\ion{O}{iii}]$\lambda$5007/[\ion{O}{ii}]$\lambda$3728 ratio (O32) derived from the sources where it was instead detected, i.e. 16.7.

We estimated the electron temperature for the low ionisation zone ($t_2$), adopting the relation $t_2^{-1} = 0.693 \, t_3^{-1} + 0.281$\footnote{Here we adopted a $t_2-t_3$ relation calibrated for an AGN continuum. We repeated the same procedure adopting a different relation suited for a star-forming continuum (\citealt{pilyugin2009electron}), but found that the average difference in the metallicity was negligible.} (see e.g. \citealt{dors2020chemical_a}). Additionally, we corrected the total oxygen abundance for the unobserved ions adopting an average ionic correction factor (ICF) equal to the mean value of those measured in the Seyfert 2 sample described in \citet{dors2020chemical_b}, that is ICF($O^{2+}$)=1.21.
For three out of four objects, this procedure yielded reasonable results with sub-solar oxygen abundances ranging between $Z=12+\log(O/H)$=7.4-8.1 and electron temperatures between 16,000 and 24,000 K. In the case of RUBIES-CEERS-55604, adopting the $A_V$ reported in \citet{kocevski2024rise}, the electron temperature inferred from the O33 ratio was unreasonably high ($T_e > 10^5$ K). At the same time, the combination of a narrow absorption bluewards of the \ion{H}{$\beta$} and the low signal to noise of the \ion{H}{$\gamma$} hampered the possibility of estimating the Balmer decrement via the \ion{H}{$\beta$}/\ion{H}{$\gamma$} ratio. For these reasons, we conservatively excluded this value from the results. The values obtained via this procedure are reported alongside other narrow emission line ratios for the low-luminosity sample in Table \ref{tbl:spec_pars}.  

\begin{figure}
	\includegraphics[width=\columnwidth]{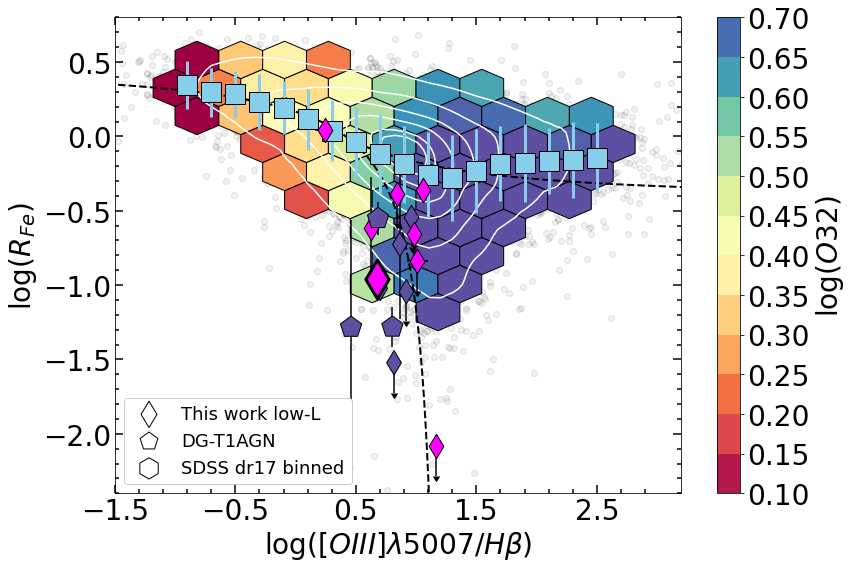}
    \caption{Distributions of the SDSS analogues and our \textit{JWST} faint AGN sample in the $\log$([\ion{O}{iii}]/\ion{H}{$\beta$}])-$\log$(\Rfe) plane with the O32 ratio colour-coded (except for the faint AGN where the [\ion{O}{ii}] was not detected). White contours show the distribution of SDSS AGN. The diamond with thicker black edges represents the value derived from the low-luminosity composite spectrum. Alongside the main anti-correlation, a secondary branch of SDSS objects detaches (the dashed lines guide the eye along the two branches), which still shares O32 values similar to those of the main trend. Notably, the \textit{JWST} low-luminosity sources follow this steeper anti-correlation trail, likely driven by low metallicity as explained in the text.}
    \label{fig:O3_Rfe}
\end{figure}

With the aim of comparing the narrow-line properties between all the \textit{JWST} low-luminosity sources and their local counterparts, in Fig. \ref{fig:O3_Rfe} we show \Rfe against the [\ion{O}{iii}]/\ion{H}{$\beta$} ratio, while in colour-code we highlight the O32 ratio for the comparison sample. 
It is well known that there are several parameters influencing the [\ion{O}{iii}]/\ion{H}{$\beta$}, namely the ionisation parameter, the metallicity, the shape of the ionising continuum, and the density (e.g. \citealt{veilleux1987spectral}). The dependence on all these parameters makes this ratio a suitable tool for interpreting observations of the nebular emission from active and inactive galaxies. In our case, the selection of a homogeneous control sample in terms of black hole mass and luminosity helps in narrowing down the parameter space. Indeed, assuming that the optical/UV region of the SED is dominated by an accretion disc, we do not expect the ionising continua to be dramatically different as the average accretion parameters are, by sample construction, close. In more quantitative terms, the difference in average $\log (M_{BH})$ and $\log(L_{5100\AA})$ between the local sample and the \textit{JWST} low-luminosity one amounts respectively to 0.15 and 0.25 dex, which is fairly close or even lower than the typical systematic uncertainty 0.3 dex and 0.2 dex. We therefore expect the shape of the continuum to be similar between these two samples as well as within the local sample itself. Similar considerations concerning how the shape of the SED should be limited by construction, apply to the diversity along the \Rfe axis in Fig.\ref{fig:O3_Rfe}. 

Additional information can be gained employing the O32 ratio, colour-coded in Fig. \ref{fig:O3_Rfe} for the SDSS control sample, which serves as a useful proxy for the ionisation parameter. In our low-luminosity sample the [\ion{O}{ii}]$\lambda \lambda$3726,3728 doublet was reliably detected, albeit unresolved, in only a handful of \textit{JWST} objects, while it was clearly measured in all the DG-T1AGN. For this reason, we only colour-code accordingly these few sources in Fig. \ref{fig:O3_Rfe}.

It is thus interesting to observe that, from the main anti-correlation trend between \Rfe and [\ion{O}{iii}]/\ion{H}{$\beta$}, which is basically a shallower version of the well-known \Rfe-[\ion{O}{iii}] anti-correlation, an even steeper anti-correlation branch detaches (see dashed lines in Fig.\ref{fig:O3_Rfe}). Notably, the \textit{JWST} low-luminosity objects sit nicely on this secondary branch. The nature of this secondary branch is not clear, but a speculative hypothesis is that this sample is a low metallicity trail. This interpretation seems favoured by the fact that the trail detaching at low values of \Rfe displays similar O32 values to those following the main trend. The same roughly holds for the faint AGN where this ratio was estimated.
Indeed, once the shape of the ionising SED and the ionisation parameter (O32) are roughly fixed, [\ion{O}{iii}]/\ion{H}{$\beta$} would increase with increasing metallicity at low metallicities and decrease with increasing metallicities at high metallicities \citep{groves2006emission, curti2016new, dors2015central}\footnote{The threshold metallicity is roughly $\sim Z/Z_{\odot}\sim 0.15$ for SF regions but can be as high as $Z/Z_{\odot}\sim 1$ for NLRs of AGN due to the harder ionizing SED.}.
Thus, we can interpret the lower left branch occupied by the low-luminosity sample as having lower metallicities compared to the SDSS sources at similar ionisation parameters, even though we do not have access to O32 for all of the low-luminosity sources.

If this scenario proved consistent, the weakness of the \Rfe in these sources could well be an effect of the reduced metal content, as this seems also the case for the three DG-T1AGN, whose metallicities are notoriously low. Still, we caution that the metallicities and ionisation parameters we infer from narrow lines only reflect the properties of NLRs. Indeed, BLRs can have chemical abundances enriched earlier than NLRs and their ionisation parameters are not necessarily connected to each other. Regardless, confirmations of metal-poor NLRs in low-luminosity sources would have room for the existence of metal-poor BLRs. In contrast, the more metal-rich NLRs in SDSS sources would likely indicate even more metal-rich BLRs. To understand more quantitatively what might be driving the difference in \Rfe, we took advantage of theoretical photoionisation models to predict \Rfe under different physical conditions in the next section.

%%%%%%%%%%%%%%%%%%%%%%%%%%%%%%%%%%%%%%%%%%%%%%%%%%%%%%%%%%%%%%%
% PARAMETER SPACE
%%%%%%%%%%%%%%%%%%%%%%%%%%%%%%%%%%%%%%%%%%%%%%%%%%%%%%%%%%%%

\subsection{Photoionisation models}
\label{sec:cloudy}

%----------------------------------------------------------------- 

\begin{table}
\setlength{\tabcolsep}{2pt}
        \centering
        \caption{Input parameters for \textsc{Cloudy} photoionisation models.}
        \label{tab:models}
        \begin{tabular}{l c}
            \hline
            \hline
            Parameter & Values \\
            \hline
            $Z/Z_\odot $ & 0.1, 0.2, 0.5, 1, 2\\
            \hline
            $\log U$& $-3.5$, $-3$, $-2.5$, $-2$, $-1.5$, $-1$ \\
            \hline
            $\log (n_{\rm H}/{\rm cm^{-3}})$& 11 (see Appendix for more values) \\
            \hline
            $\log (N_{\rm H}/{\rm cm^{-2}})$ & 24, 25 \\
            \hline
            Microturbulence & $v_{\rm turb} = 100~{\rm km~s^{-1}}$\\
            \hline
            AGN SED & $T_{\rm BB}=10^6~{\rm K};~\alpha _{\rm ox}=-1.4, -5$ \\
            \hline
            Solar abundance set & \citet{grevesse_solar_2010} abundance set\\
            \hline
            %N/O (C/O) vs. O/H & \citet{groves_no_2004} relation\\
            %\hline
            %He/H vs. O/H & \citet{dopita_heh_2000} relation\\
            %\hline
            Dust & No dust\\
            \hline
            \Feii atomic data& \citet{bautista_fe2data_2015},\\
             &\citet{tayal_fe2data_2018},\\
             & \citet{smyth_fe2data_2019}\\
            \hline
        \end{tabular}
\end{table}

%----------------------------------------------------------------- 

\begin{figure*}
    \includegraphics[width=\columnwidth]{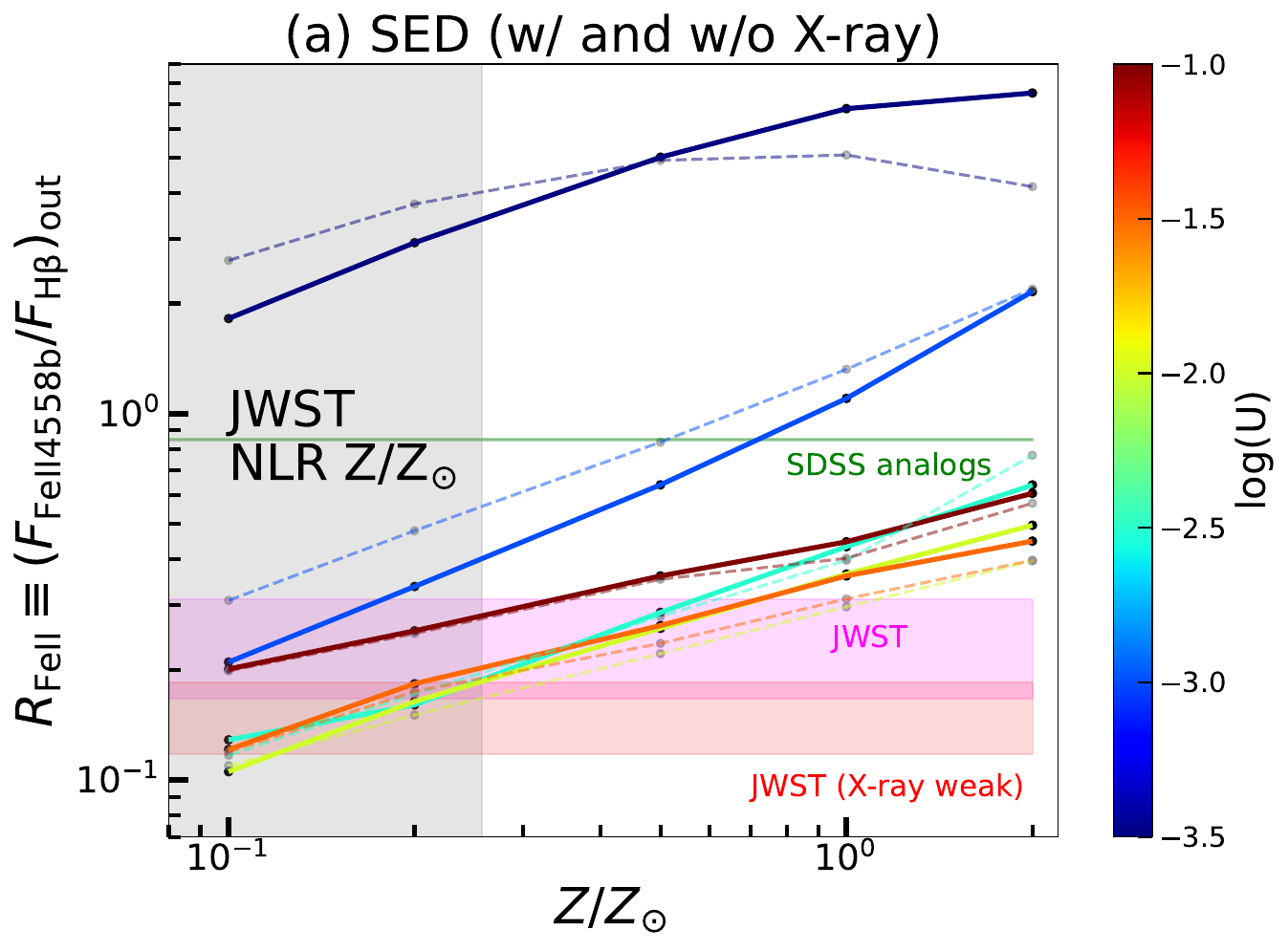}
    \includegraphics[width=\columnwidth]{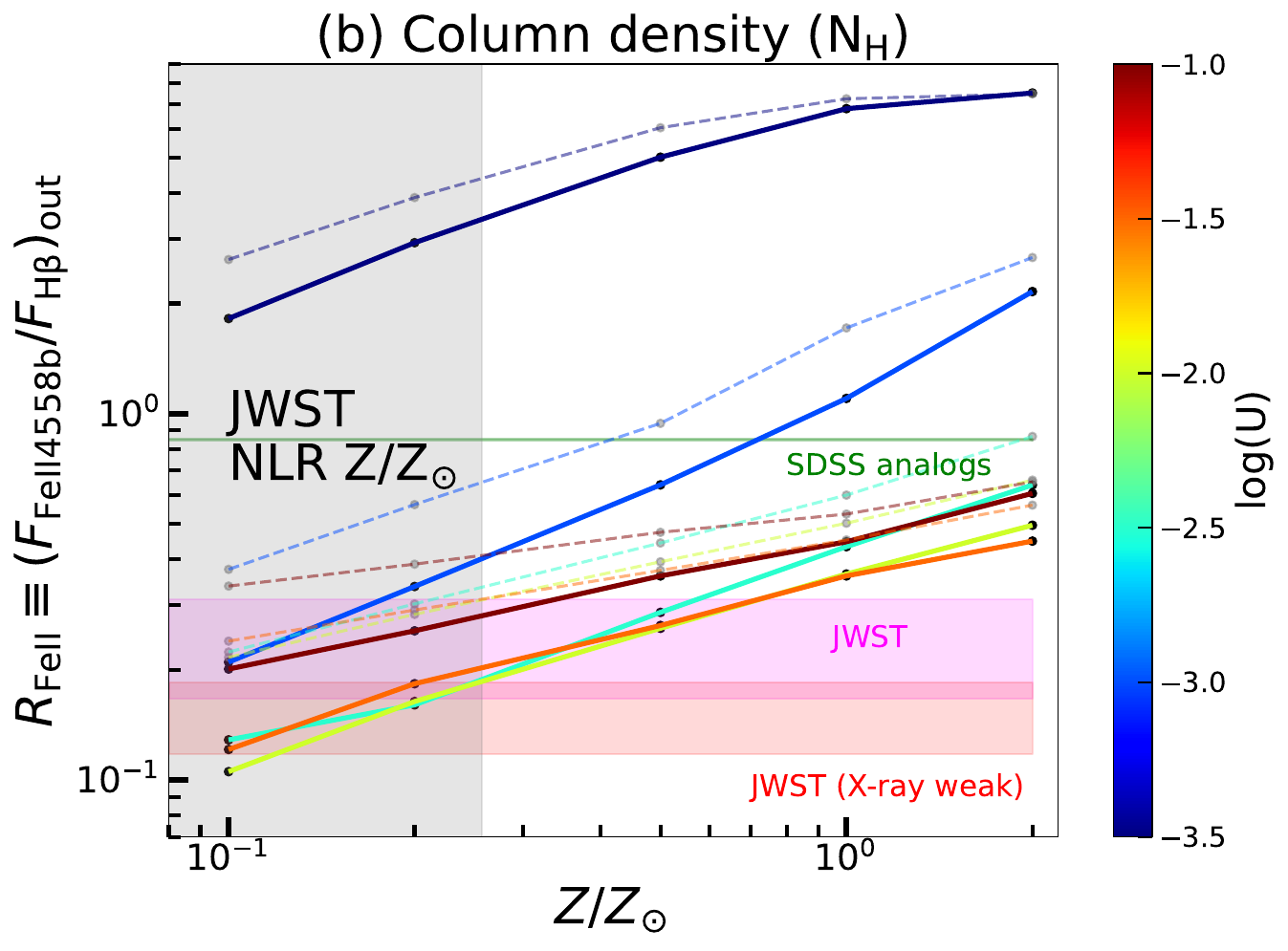}
    \caption{Comparisons between \Rfe predicted by photoionisation models with different input parameters and \Rfe measured in our sample. In each panel, the green shaded band indicates the mean \Rfe and its $1\sigma$ uncertainty found in the SDSS sample that matches the \textit{JWST} sample in terms of \Hb luminosities and FWHM; the magenta shaded region corresponds to the value found in the \textit{JWST} sample; the red shaded region corresponds to the value found in the \textit{JWST} sample excluding sources with X-ray detections. The grey shaded region indicates the plausible range of metallicities in the NLRs of the \textit{JWST} sources based on the $T_{\rm e}$ method for a subset of the \textit{JWST} sample. Photoionsation models predictions are plotted as a function of the metallicity, Z/Z$_{\odot}$, and are color coded by the ionisation parameter.
    \textit{Left:} solid lines and dashed lines correspond to models with and without a hard X-ray component representing the hot corona contribution to the SED, respectively,
    \textit{Right:} solid lines and dashed lines correspond to models having $N_{\rm H} = 10^{24}~{\rm cm^{-2}}$ and $N_{\rm H} = 10^{25}~{\rm cm^{-2}}$, respectively.
    }
    \label{fig:cloudy_Rfe}
\end{figure*}

To better understand the difference in \Rfe between the low-$z$ SDSS sample and the high-$z$ \textit{JWST} sample, we use theoretical photoionisation models computed with \textsc{Cloudy} \citep[v17.03,][]{cloudy17}.
During the computation of the models, we turn on all levels available for $\rm Fe^+$ with atomic data sets implemented by \citet{sarkar2021improved}, from \citet{bautista_fe2data_2015}, \citet{tayal_fe2data_2018}, and \citet{smyth_fe2data_2019}.
The theoretical computation of the strength of \Feii has been investigated by many previous works, which showed that \Rfe broadly depends on parameters such as the Fe abundance, the gas volume density, the gas column density, the ionising photon flux, the shape of the ionizing SED, and the microturbulent velocity \citep[e.g.,][]{baldwin2004origin,ferland_fe2_2009,panda2018modeling,temple2020fe,sarkar2021improved}.
For the purpose of this work, we make and justify the following assumptions and simplifications during photoionisation modelling, with the model parameters summarised in Table~\ref{tab:models}.

First, we restrict the comparison between the low-$z$ and high-$z$ samples to a region within the observed parameter space with similar broad \Hb luminosities and FWHM (see Fig. \ref{fig:sel}). Based on this selection criterion, we assume both samples of AGN to share similar accretion properties and thus do not differ significantly in their ionising SED produced by the accretion discs.
Specifically, we assume the shape of the SED to follow the functional form 
\begin{equation}
    F_\nu = \nu ^{\alpha _{\rm uv}}{\rm exp}(-h\nu /kT_{\rm BB}){\rm exp}(0.01~{\rm Ryd}/h\nu) + a\nu ^{\alpha _{\rm x}},
\end{equation}
where by default we set the temperature of the cutoff of the ``Big Blue Bump'' component to $T_{\rm BB} = 10^{6}$ K, the UV-to-X-ray slope to $\alpha _{\rm ox} = -1.4$ (by adjusting $a$ in the equation above), the UV slope to $\alpha _{\rm uv}=-0.5$, and the X-ray slope to $\alpha _{\rm x} = -1.0$.
It is worth noting that this assumption might not hold over the full energy range of the SED, as the majority of the \textit{JWST} identified AGN in our sample is clearly X-ray weak.
Recently, \citet{maiolino2024jwst} suggested two potential reasons for the lack of hard X-ray detections in early AGN, one being the intrinsic lack of the hard X-ray component due to the missing hot corona, the other being the presence of Compton-thick and dust-free clouds within the BLRs.
To take into account these effects, we perform two additional tests by computing models with suppressed X-ray components and models with different hydrogen column densities.

Second, we assume a fixed gas density and a fixed microturbulence velocity.
Specifically, we adopted a hydrogen density of $n_{\rm H} = 10^{11}~{\rm cm^{-3}}$, a value encompassed by those explored in many previous \Feii modelling works, typical for BLR regions (e.g., \citealt{joly1987formation, netzer_agnline_1990, collin2000fe, sigut2003predicted,baldwin2004origin,ferland_fe2_2009, panda2018modeling,sarkar2021improved}).
While there is evidence that the ISM density in galaxies is gradually increasing with redshift, there is neither observational evidence nor theoretical expectation that a similar evolution should exist in the BLR surrounding supermassive black holes. 
Systematic studies about the evolution of the broad line strength with redshift point in the direction of no evolution with cosmic time (\citealt{croom2002correlation, stepney2023no}).
The same consideration applies to the ionisation parameter, which is observed to increase with redshift in the ISM, but its redshift evolution within BLRs is not clear and its value strongly depends on the geometry.
We further discuss the effect of the ionisation parameter together with that of the metallicity later in this section.
Regardless, we check the effect of the density variation by computing models with $n_{\rm H} = 10^{10}~{\rm cm^{-3}}$ and $n_{\rm H} = 10^{12}~{\rm cm^{-3}}$, respectively.
In general, at lower density, \Rfe is enhanced at low ionisation parameters but suppressed at high ionisation parameters.
At the high density, the opposite effect of the ionisation parameter on \Rfe is seen.
This can be understood as the fact that \Feii is optimally emitted by clouds with certain combinations of densities and ionising fluxes \citep[e.g.,][]{baldwin2004origin}, and the latter is simply the product of the ionisation parameter and the density.
In Appendix~\ref{appendix:varyden}, we further discuss the effect of density variations.

The microturbulence is usually included in the photoionisation modelling of BLRs as a source for producing large Doppler broadening of line profiles within the photoionisation mean free path for line productions.
The microturbulence strengthens the continuum pumping and line fluorescence that produce \Feii emission, and reduces the optical depths of \Feii lines (e.g. \citealt{netzer1983broad,baldwin2004origin}).
Although the origin of the microturbulence in BLRs remains debated (see Sec. 8.1 in \citealt{baldwin2004origin} for a review of possible mechanisms), it is generally agreed among the previous modelling works that a microturbulent velocity of at least $v_{\rm turb} = 100~{\rm km~s^{-1}}$ is needed to correctly predict \Feii emission (\citealt{panda2019cloudy}), which is also what we adopted in our models. One potential way to connect the microturbulence to the other peculiar features of this sample, in particular the X-ray weakness, is to consider the magneto-hydrodynamic (MHD) wave explanation, where the microturbulence comes from non-dissipative waves generated by magnetic fields \citep{bottorff_mhd_2000}. If the hot corona producing the hard X-ray emission ($E\gtrsim$2 keV) is missing in these low-luminosity AGN due to the absence of the magnetic field lifting the corona from the accretion disk, it might be reasonable that the microturbulence is also suppressed and even vanishing in these early AGN, leading to reduced emergent \Feii emission. The issue with this explanation is that the reason for the vanishing magnetic fields is not clear and it is also not clear whether the microturbulence should rise from the MHD waves.

We note that it is still possible that the density and microturbulence play a role in producing the difference in \Rfe we observed. Our choice of not varying these parameters is mainly motivated by the fact that their redshift evolution is neither expected nor observed. The main parameters we inspect are the metallicity (or, more precisely, the iron abundance), the power-law X-ray component of the SED, and the hydrogen column density. Although previous works on BLR metallicities of bright quasars suggest fast chemical evolution within BLRs \citep[e.g.,][]{dietrich2003fe,maiolino2003early,wang2022metallicity}, our sample mainly consists of lower luminosity Type 1 AGN and it is not unreasonable to speculate they are more metal-poor compared to the local AGN. Checking the effect of the X-ray component and the hydrogen column density, as we have explained, is motivated by the fact that the \textit{JWST} identified Type 1 AGN are mostly X-ray weak. One potential explanation is the lack of hot coronae in these AGN, meaning the absence of the power-law X-ray component in their SEDs, an occurrence verified for instance in \citet{ricci2020destruction} to explain the drastic transformation of the X-ray properties following a changing-look event. 
The other potential explanation is the presence of Compton-thick ($N_{\rm H} \gtrsim 1.5\times 10^{24}~{\rm cm^{-2}}$) and dust-free clouds within their BLRs. While such clouds might not be uncommon for line-emitting clouds within BLRs \citep{netzer_Nh_2009}, to significantly obscure X-ray emission, they need to have a large covering fraction near unity. This might also explain the high EWs of broad H$\alpha$ in early AGN (\citealt{maiolino2023jades, wang2024rubies}) and could be caused by the metal-poor environments and the consequent ineffective removal of dense gas through radiation pressure.
Regardless, we compute additional models with $N_{\rm H} = 10^{25}~{\rm cm^{-2}}$ to check the potential effect with a further increase in the column densities of BLR clouds (compared to $N_{\rm H} = 10^{24}~{\rm cm^{-2}}$ by default).

In Fig.~\ref{fig:cloudy_Rfe}, we compare our \textsc{Cloudy} model predictions with \Rfe measured in our sample.
Specifically, from our sample, we plot the mean \Rfe of the whole \textit{JWST} sample, the X-ray weak \textit{JWST} sample, and the SDSS analogues matched in \Hb luminosities and FWHM, together with the corresponding $1\sigma$ uncertainties.
The mean \Rfe of the SDSS sample is roughly 3.5 times that of the whole \textit{JWST} sample, which is roughly 1.6 times that of the X-ray weak \textit{JWST} sample. We plot the model predicted \Rfe as a function of the metallicity and the ionisation parameter.
At fixed ionisation parameters, \Rfe generally increases with increasing metallicities.
Although \Feii transitions are important cooling channels in the partially ionised zones within the BLR clouds, meaning their strengths are largely determined by the heating and cooling equilibrium rather than the Fe abundance, this effect is stronger for the UV \Feii transitions, while the optical \Feii transitions have a stronger metallicity dependence \citep{shields2010}.
At fixed metallicity, on the other hand, \Rfe does not exhibit a strong dependence on the ionisation parameter for $\log U > -2.5$ and only starts to significantly increase with decreasing ionisation parameters for $\log U < -2.5$.
At low ionisation parameters, the intensity of \Hb drops significantly due to the drop in the ionisation rate. In contrast, due to the low ionisation potential of Fe, the intensity of \Feii from the partially ionised zone decreases more gradually than the \Hb.
%{\color{red} Look at EWs as well?}
If we only focus on the high ionisation parameter branch, the mean \Rfe of the SDSS sample is consistent with $Z/Z_{\odot} \gtrsim 2$, and the mean \Rfe of the \textit{JWST} sample can be explained by $0.1 \lesssim Z/Z_{\odot} \lesssim 0.5$.
In comparison, we mark the potential range of the NLR metallicities for the \textit{JWST} sample with the shaded region.
We note that the BLR metallicities could be well above the NLR metallicities, and the abundance ratio of Fe/O could also be different. Regardless, from these models, the metallicity difference appears as a plausible explanation for the much lower \Rfe in the \textit{JWST} sample.

On the left panel of Fig.~\ref{fig:cloudy_Rfe}, we simulate the effect of an intrinsic X-ray weakness by virtually removing the corona. We achieved this by computing a set of models with $\alpha _{\rm ox} = -5$, which effectively makes the power-law X-ray component contributed by the hot corona negligible compared to the emission from the accretion disc.
The resulting models are plotted as dashed lines, which do not predict very different \Rfe compared to our fiducial models with $\alpha _{\rm ox} = -1.4$ at given metallicities and ionisation parameters. This is expected as the hard X-ray component should generally contribute little to the ionisation as well as heating of the gas, and consequently \Rfe.
In addition, compared to the accretion disc emission, the hot corona contributes little to the total soft X-ray emission that is responsible for creating the partially ionised zone where \Feii originates from.
We emphasise again that we assume no significant difference in the shape of the accretion disc emission between the SDSS analogues and low-luminosity sources, given their accretion parameters are empirically constrained to be similar by their broad \Hb luminosities and FWHM. On the right panel of Fig.~\ref{fig:cloudy_Rfe}, we check the effect of thicker BLR clouds with larger column densities reaching $N_{\rm H} = 10^{25}~{\rm cm^{-2}}$.
As expected, increasing the column density increases \Rfe. This is because \Hb is mainly produced by the ionised layer of the clouds, whereas \Feii becomes the major coolant at large cloud depths, and thus increasing the cloud depths tends to boost \Rfe. However, this trend is opposite to what is observed, that is, a decrease in \Rfe at early times.

To summarise, neither the hot corona emission nor the column density seems to cause the observed difference in \Rfe.
The metallicity is likely the most important factor governing the \Rfe in our sample, consistent with our interpretations based on \Rfe versus narrow-line ratios in Fig.~\ref{fig:O3_Rfe}. These analyses point towards the picture where, unlike luminous quasars at early times, the low-luminosity AGN in our sample have less chemically evolved BLRs or, at least, less Fe-enriched BLRs.
We further discuss the implications for the chemical evolution of these systems in Section~\ref{sec:discussion}.

\subsection{An X-ray perspective}
\label{sec:Xray_perspective}
The low luminosity AGN population discovered by \textit{JWST} is prevalently undetected in the X-ray surveys (\citealt{yue2024stacking, maiolino2024jwst}) with the detection efficiency being below the percent level (see e.g. \citealt{kocevski2024rise}). Our sample makes no exception, with all (but two) of the low luminosity sources being undetected in the X-rays in their respective fields. This also applies for the three newly discovered RUBIES objects where no X-ray counterpart was found in the 800 ks of the Chandra Aegis-X Deep survey catalogue (\citealt{nandra2015aegis}). The only two non quasar sources which escape this picture are XID-2028 and JADES-209777, both of them reliably X-ray detected as reported in \citet{brusa2010xmm} and Juod{\v{z}}balis et al. in prep. Intriguingly, these sources are the objects with the highest \Rfe below the quasar luminosity regime (respectively \Rfe$=1.10$ and \Rfe$=0.44$), and in line or even above the \Rfe expectations for the local sources in the corresponding region of the parameter space (see the colour-code in Fig. \ref{fig:parspace_rfe}). Although it is certainly risky to draw conclusions based on such a small-number statistics, it is interesting to notice that the only two X-ray detected sources of the sample are also those on the high end of the \Rfe distribution for the \textit{JWST} low-luminosity objects. 

From a theoretical perspective, it is not straightforward to link the X-ray properties with the \Rfe strength. As we showed in Sec. \ref{sec:cloudy}, the intensity of the optical iron does not strongly correlate with the hard X-rays component. This is due to the fact that the \ion{Fe}{ii} is mostly produced at optical depths where the heating and the ionisation are due to lower energy soft X-ray photons rather than hard X-ray photons. Indeed, from an observational standpoint, \Rfe does instead seem to positively correlate with the so-called soft excess, i.e. the low-energy ($\lesssim 0.5$ keV) extrapolation of the hard X-ray power law (\citealt{wilkes1987optical,wang1996, shastri1993quasar}). This broadly agrees with the findings on narrow-line Seyfert 1 galaxies, whose X-ray spectra on average steep, being generally stronger optical iron emitters. 
Here we mention two possible scenarios to link X-ray and \Rfe weaknesses. In the first case, the accretion parameters characterising the local and the high-$z$ sources are systematically disparate because of different $M_{BH}$ and $L_{bol}$ calibrations between these two populations. Therefore, these classes of sources have intrinsically distinct soft X-ray SEDs that lead to the observed \Rfe difference. Secondly, it is possible that the low metallicity of the BLR clouds causes both the X-ray and the \Rfe weakness. We already showed in Sec. \ref{sec:cloudy} that the \Rfe observed in the faint \textit{JWST} AGN are consistent with metal-poor BLR. Qualitatively, we can speculate that this effect might also lead to the BLR clouds lingering close to the innermost AGN engine because of the low effective radiation pressure. Therefore, the reduced metal content devoids the BLR clouds of the metal line transitions which help reaching the dynamical equilibrium.

We also highlight that both these X-ray detected sources have substantially lower redshifts than the bulk of the low luminosity sample ($\langle z \rangle$=6.34), in particular $z=1.59$ for XID-2028 and $z=3.71$ for JADES-209777. The only other object at a somewhat lower redshift from the bulk of the redshift distribution of the low luminosity sample is JADES-028074 (\citealt{juodvzbalis2024jades}) at $z$=2.26. This object is of utmost interest as it encapsulates some of the typicalities of the faint \textit{JWST} AGN, such as the X-ray (and Radio) weakness, the lack of ionised outflows, and hereby we also add the \Rfe weakness, at the same time showing correlated absorptions in the \Hb, \ion{H}{$\alpha$} and \ion{He}{i}. Within an evolutionary framework, sources akin to JADES-028074, might be experiencing the transitioning phase between an AGN with the characteristics of the faint \textit{JWST} AGN and those of Seyfert 1 AGN observed at lower $z$ exemplified by JADES-209777 and XID-2028. In a wider perspective, the significant fraction of JWST-detected AGN with absorption features in the broad lines (\citealt{matthee2024little, kocevski2024rise, wang2024rubies, maiolino2024jwst, juodvzbalis2024jades}) could be experiencing the ``blow-out'' phase bridging between the obscured and unobscured stages of the AGN cycle (\citealt{hickox2018obscured}). Yet, the outflow velocities reported in this case are significantly lower than those seen in the nuclear region of quasars measured from broad absorption lines or in X-ray spectra (a few hundred versus tens of thousands km s$^{-1}$).

%##########################################################################
\section{Discussion}
\label{sec:discussion}

The first years of \textit{JWST} observations have revolutionised our understanding of the high-redshift Universe, in particular by sheding a new light on an enigmatic population of faint AGN. These sources have been discovered to generally share a set of observational characteristics that distinguish them from local counterparts of similar luminosity and hosting black holes of comparable mass. Examples of these observational features are the weakness of the ionised outflows, routinely observed in low-$z$ sources as blue wings in the [\ion{O}{iii}] profile, the incidence of absorption features in Balmer lines, and their remarkable X-ray and radio weakness. Hereby we also demonstrated the weakness of their \Rfe ratios when compared to a sample of local Type 1 AGN with similar \ion{H}{$\beta$} properties.

The strength of the \Rfe ratio is influenced by multiple factors \citep[see e.g.,][]{netzer1983broad,joly1987formation,verner_fe2_1999,baldwin2004origin}, such as the shape and the intensity of the incident continuum, the cloud density, the column density, the metallicity, and the microturbulence, which concur to produce the optical spectral diversity of Type 1 AGN (e.g. \citealt{panda2019cloudy}). Although we have no means to directly infer the density and the turbulent velocity, we could narrow down the parameter space in terms of the incident SED, by selecting a local Type 1 control sample with comparable $\rm M_{BH}$ and $\rm L_{H\beta,br}$ (and likely $\rm L_{bol}$). This under the hypothesis that the shape of the accretion disc SED only depends on $\rm M_{BH}$ and $\rm L_{bol}$ with the exception of the hard X-rays that, as we showed, are not crucial for the \Feii production. Our main finding is that, on average, the \textit{JWST} low-luminosity sources exhibit significantly (p-value$\lesssim 10^{-6}$) lower \Rfe than their local counterparts.

The same, instead, does not apply to the more luminous objects of the sample, which closely resemble their lower redshift counterparts. The similarity in terms of continuum shape and emission lines -albeit some known trends with the luminosity, e.g. the Baldwin effect (\citealt{baldwin1977})- is not surprising, since high-redshift quasars have often been observed to match the spectral properties of typical sources at $z\lesssim$1 (e.g. \citealt{kuhn2001search, mortlock2011luminous, hao2013spectral, banados2018, fan2023quasars, trefoloni2024quasars}). This implies that, at least for the bright end of the AGN distribution, no signs of clear evolution in the spectral properties can be observed up to $z\sim 7$. 

Similar considerations have been drawn from the chemical enrichment of the BLR of luminous quasars which, parametrised for instance with the UV \ion{Fe}{ii}/\ion{Mg}{ii} ratio, does not even show any evolutionary trends up to redshift $\sim$7.5 (see e.g. \citealt{dietrich2003fe, mazzucchelli2017physical,  sameshima2020mg, wang2022metallicity, trefoloni2023most, jiang2024no}), although the dependence on the Fe and Mg abundances of this ratio is quite weak (\citealt{sarkar2021improved}). In this context, it is interesting to note that also a detailed photoionisation modelling of the shallow UV \Feii bump in the faint high-redshift AGN GN-z11 (\citealt{maiolino2024small}) led to infer a sub-solar BLR metallicity (\citealt{ji2024jades}).

The most straightforward explanation for this lack of evolution resides in the fact that quasars are expected (e.g. \citealt{costa2014environment}) and observed (e.g. \citealt{cantalupo2014cosmic, mignoli2020web, overzier2022conditions}) to reside in overdense regions of the cosmic web, and therefore likely represent the most advanced stage of the chemical evolution at all the cosmic times. This effect has often been epitomised in the so-called luminosity-metallicity relation ($L-Z$ relation; \citealt{hamann1993chemical, hamann1999elemental, dietrich2003quasar}). In these spots, the first generation of low- and intermediate-mass evolved stars might have been able to promptly produce heavy elements even at $z\gtrsim6$, following the rapid enrichment of $\alpha$-elements owing to type II supernovae and leading to BLR metallicities ranging from $Z\sim Z_{\odot}$ up to extreme values.

All the luminous sources in our sample are X-ray detected (or observations targeting them are undergoing), while the low-luminosity AGN are mostly not (12/14). Notwithstanding the nature of the X-ray weakness (intrinsic or rather due to absorption) in these sources has not been understood, the overwhelming fraction of X-ray weak objects in this redshift and luminosity regime calls for an explanation. Although several mechanisms involving, for instance, a systematic increase in the ISM density have been proposed to explain the increasing X-ray obscured fraction (e.g. \citealt{gilli2022supermassive, alonso2024probing}), such a large amount seems hardly reconcilable with those observed at lower redshift. As a term of comparison, the X-ray weak fraction ranges between 2\% in optically selected samples (e.g. \citealt{Gibson2009}) to $\sim$20\% in samples of broad absorption line objects such as that in \citealt{liu2018frequency}, reaching as high as $\sim$25\% for the most luminous blue quasars at z$\sim$3 (\citealt{nardini2019}). Yet, even the highest of these X-ray weak fractions is far below the non-detection rate of the faint \textit{JWST} AGN.

An obvious question is how these sources relate to the local population of AGN and how the transition between these apparently different classes of sources might take place. Some clues about a tentative evolutionary path bridging between the population of faint high redshift and local AGN, could be embedded in the fraction of sources displaying blueshifted Balmer absorption features, which the latest compilations report to be at least 10\% (see Section 7 in \citealt{juodvzbalis2024jades} for a more complete discussion). In our case, we detect significant absorption features in 2/14 low-luminosity AGN. This fraction is significantly larger than the 0.1\% found in the SDSS Type 1 compilation at lower redshift (\citealt{juodvzbalis2024jades}). Yet, this value could be somewhat underestimated, given that the typical SDSS resolution is lower than the high-resolution \textit{JWST} grating, that the SDSS spectra have generally lower SNR, and that a systematic search for Balmer absorptions in a large sample of SDSS Type 1 AGN has not yet been undertaken.

Balmer lines absorption requires peculiar conditions of the gas along the line of sight, in particular a substantial amount of mostly neutral gas with a significant fraction of hydrogen pumped to the n = 2 level. Due to the short lifetime of H(n = 2), it is quite challenging to have prominent absorption in hydrogen lines without the presence of high-density gas with column densities of the order of $N_{\rm H}\sim 10^{21-23}$ cm$^{-2}$ (the interested reader may consult Sec 5.2 in \citealt{juodvzbalis2024jades} for a more complete discussion of the gas properties required for Balmer absorption). The large column densities required to produce Balmer absorption, combined with the recent findings of larger \ion{H}{$\alpha$} equivalent widths in \textit{JWST} low-luminosity objects (\citealt{maiolino2024jwst, wang2024rubies}), could fit in the scenario where the BLR has both a high column density and a high covering factor. These factors, could also be key in producing the extremely high fraction of X-ray non-detections, by means of dust-free absorption.

In Sec. \ref{sec:cloudy} we explored which physical parameters could be the most conducive to driving the observed \Rfe weakness, by means of photoionisation modelling. In doing so, we adopted observationally educated guesses, while, at the same time, narrowing down the parameter space of the control sample. We employed the strong observational constraints about the X-ray emission in the \textit{JWST} low-luminosity AGN to test whether an intrinsic lack of the hard X-ray coronal emission could play any role in producing the low \Rfe observed within our sample, but we found none. This is in line with several other works reporting the irrelevance or even the anti-correlation between the strength of the hard X-ray component (parametrised via the spectral index) and the \Rfe ratio (see e.g. \citealt{shastri1993quasar, wang1996, laor1997soft, wilkes1999investigation, shenho2014}). As a consequence, we argue that the X-ray weakness cannot account for the observed \Rfe weakness. 
At the same time, we also excluded systematic differences in the ionising SED as the drivers of the observed difference between the \textit{JWST} low-luminosity and the local AGN, under the hypothesis that the spectra of accretion discs with similar $\rm M_{BH}$ and $\rm L_{bol}$ do not vary greatly with redshift. Such non-evolution seems to hold, at least, for the UV/optical regions of the accretion disc SED in quasars which show a remarkable similarity between low and high-$z$ (e.g. \citealt{trefoloni2024quasars}).
Although there is evidence for an increase of both the ISM ionising parameter (see e.g. \citealt{reddy2023impact} and references therein) and electron density (see e.g. \citealt{isobe2023redshift}), little however is known about the redshift evolution of these properties in the BLR. Even less is known about the causes for the microturbulence invoked to give the higher \Feii multiplets access to a greater range of exciting continuum photons, thus increasing the \Feii equivalent width (\citealt{netzer1983broad, baldwin2004origin, bruhweiler2008modeling}). Under the hypothesis that these quantities are not systematically different between the low-luminosity \textit{JWST} and the local AGN, we showed that the most likely responsible for the low \Rfe measured is the BLR metallicity, as already pointed out for local sources in \citet{floris2024chemical}, although we cannot directly probe it in our sample.

The low metallicity could be the keystone to explain the exotic properties in both the NLR and BLR of \textit{JWST} low-luminosity objects. Indeed, as the low metal (or at least Fe) content seems the most viable explanation for the weakness of the \Rfe, the reduced metal abundance in the BLR gas could, at least qualitatively, imply significantly fewer line transitions to provide radiation pressure support. This in turn, would translate into a closer, and hence more covering, BLR. A low-metallicity framework also goes in the direction of explaining the scarcity of prominent [\ion{O}{iii}] outflows, which struggle to develop on galactic scales due to the reduced radiation pressure, as already pointed out in \citealt{maiolino2023jades}. If this were the case, the scarcity of metals would also naturally explain the several similarities between the faint \textit{JWST} AGN and those observed in local metal poor dwarf galaxies. 

Valuable insights on the physical properties of these high-redshift sources could be gathered, if the analogy between the faint \textit{JWST} AGN and those in metal poor dwarf galaxies actually holds.
Recently, \citet{doan2024local} proposed that SDSS J1201+0211, a compact metal-poor dwarf galaxy, possibly hosting an AGN, could represent a local analogue to the primordial galaxies. Here the authors focus on the locus occupied by this object in the BPT diagram, which is also populated by low-metallicity star forming (dwarf) galaxies, as well as high-z AGNs recently discovered by \textit{JWST} (e.g. \citealt{maiolino2023jades, harikane2023jwst, ubler2023ga}). Although a more quantitative and systematic assessment of the similarities between the faint high-redshift AGN population and that in metal poor dwarf galaxies has not yet been undertaken, some commonalities are apparent. For instance, weak ionised outflows (if any), mostly detectable in the [\ion{O}{iii}], the X-ray weakness and the scarcity of heavy elements characterise both these classes of objects (although we report a faint outflow component in J1025+1402). Hereby, we also highlight the \Rfe weakness revealed by the analysis on the three DG-T1AGN with broad \Hb, which our analysis proved consistent with a reduced iron content. Additionally, we also remark, as a further similarity, the high incidence of Balmer absorption features encountered in the dwarf galaxies hosting broad line AGN. Indeed, one out of the four DG-T1AGN described in \citet{burke2021agn} (J1025+1402), also included in our sample, shows a clear absorption in the \ion{H}{$\alpha$} profile. Moreover, we also report a significant absorption in the \ion{H}{$\beta$} profile of SBS\_0335-052E ($\Delta$BIC=44), thus yielding a 40\% detection rate, although on a fairly small sample.

To summarise, evidence is growing in favour of metallicity as being the ultimate responsible for the observed differences between the local and the high-redshift AGN populations. Yet, additional observations and modelling will prove key in constraining the physical properties of this newly discovered, intriguing class of AGN.

\section{Conclusions}
\label{sec:conclusions}
We assembled a sample of high-redshift AGN observed by \textit{JWST} spanning a wide range in terms of luminosity and black hole mass with the aim of characterising the strength of the \Rfe. To this end, we performed a detailed spectroscopic analysis and selected control subsamples at lower redshift to compare their properties and assess possible differences. We remark that our control samples are selected to match the \textit{JWST} AGN in both \Hb FWHM and luminosity, which are the direct emission line observables that best capture the black hole properties and accretion parameters. Additionally, we also included as a separate comparison sample three metal-poor dwarf galaxies hosting low-luminosity AGN as tentative super local counterparts. Here we summarise our findings:

\begin{itemize}  
    \item The strength in the ratio between optical \Feii and the \Hb equivalent widths (\Rfe) in our \textit{JWST} low-luminosity AGN is significantly lower (p-value $<10^{-6}$) than in their local SDSS counterparts with similar BH accretion parameters. This \Rfe weakness is not compatible with the expectations from known sets of correlations such as the 4DE1.
    \item The strength in the \Rfe of the high-luminosity (quasar) subsample is instead consistent with the expectations derived in a comparison sample of quasars between $1.5 < z < 3.5$.
    \item Our photoionisation modelling suggests the weakness of the \Rfe in faint AGN is consistent with low metal content in their BLRs. In contrast, high luminosity sources likely reached the chemical maturity observed at early cosmic times.
    \item The faint AGN hosted in local metal poor dwarf galaxies (DG-T1AGN) also exhibit strikingly low \Rfe. Also in this case, the driver for the observed weakness is expected to be the lack of metals. 
    \item In all the parameter spaces explored, faint AGN in dwarf metal-poor galaxies (DG-T1AGN) occupy the same locus as the low-luminosity \textit{JWST} AGN. Additionally, they also seem to share other characteristics such as X-ray weakness, faint or absent ionised outflows and a tentative high incidence of absorption features in Balmer lines. These findings go in the direction of DG-T1AGN being close local analogues to the faint high-redshift AGN population, barring a potential difference in their host galaxy properties to be verified by follow-up studies.
\end{itemize}

The launch of the \textit{JWST} space telescope three years ago opened a new era for astronomy, astrophysics and cosmology. It made possible, for the first time, not only to observe the brightest objects shining only a few hundred Myr after the Big Bang, but even to start collecting a census of the elusive populations of faint sources already in place at that time.

The results elaborated here point in the direction that the realm of AGN shows hints of bimodality. If luminous quasars at redshift $\gtrsim 6$ seem to exhibit features broadly consistent with those observed at lower redshifts, the same does not apply to faint sources which display quite peculiar features. A comprehensive model capable of explaining such exotic behaviours (such as the X-ray and \Rfe weakness, the faint ionised outflows, the incidence of Balmer absorption) is still far from being designed. Yet, pieces of evidence in favour of the metallicity playing a key role are being collected.

Obtaining new (possibly large) panchromatic datasets will be key to consolidate our knowledge on a more solid base, as well as to infer many new details still overlooked. To this end, a crucial role will be played by the new generation of optical and infrared ground based telescopes such as Extremely Large Telescope (ELT; \citealt{gilmozzi2007european}), the Giant Magellan Telescope (GMT; \citealt{sanders07}) and the Thirty Meter Telescope (TMT; \citealt{skidmore2015thirty}). This new class of observing facilities will allow us to investigate the Universe up to the dark ages with unprecedented sensitivity. 
Another viable way to strengthen our knowledge about the faint \textit{JWST} AGN population is a more thorough assessment of their similarity with AGN in local dwarf metal poor galaxies. If these sources were confirmed to be reliable very local counterparts of the high-redshift population, this would enable us to get a deeper understanding of the high-redshift Universe with unparalleled sensitivity and spatial resolution.

\begin{acknowledgements}
The authors acknowledge Gary Ferland for the fruitful discussion and A. de Graaf and the RUBIES team for the invaluable work dedicated to the RUBIES survey.
BT gratefully thanks L. Furtak for providing the reduced spectrum of Abell2744-QSO1, F. Loiacono for that of J0100+2802, V. Kokorev for that of UNCOVER-20466, M. Marshall for those of DELS J0411-0907 and VDES J0020-3653, A. Eilers for that of PJ308-21, G. Cresci for that of XID-2028 and Y.I. Izotov for that of SBS\_0335-052E. XJ, RM, FDE, GM and JW acknowledge support by the Science and Technology Facilities Council (STFC), by the ERC through Advanced Grant 695671 “QUENCH”, and by the UKRI Frontier Research grant RISEandFALL. RM also acknowledges funding from a research professorship from the Royal Society. H\"U gratefully acknowledges support by the Isaac Newton Trust and by the Kavli Foundation through a Newton-Kavli Junior Fellowship.

This work made use of the Python packages: Astropy \citealt{robitaille2013astropy}) and \textit{dust\_extinction}, numpy (\citealt{harris2020array}), Matplotlib (\citealt{Hunter:2007}), pandas \citealt{reback2020pandas}, SciPy (\citealt{2020SciPy-NMeth}). 
This work is based on observations made with the NASA/ESA/CSA James Webb Space Telescope. This work made use of spectra from the SDSS survey, funding for the Sloan Digital Sky Survey IV has been provided by the Alfred P. Sloan Foundation, the U.S. Department of Energy Office of Science, and the Participating Institutions.
\end{acknowledgements}

%%%%%%%%%%%%%%%%%%%%%%%%%%%%%%%%%%%%%%%%%%%%%%%%%%
\section*{Data Availability}

All data used in this study are publicly available and will be provided upon reasonable request to the author. Part of the \textit{JWST} data presented in this paper were obtained from the DAWN \textit{JWST} Archive (DJA) repository at \url{https://dawn-cph.github.io/dja/}.

\bibliographystyle{aa} 
\bibliography{bibl}

\begin{appendix}

\twocolumn

\section{Systematic \texorpdfstring{\Rfe}{Rfe} differences due to the fitting technique}
\label{app:Rfe_sys_diff}
\begin{figure}[h!]
	\includegraphics[width=\columnwidth]{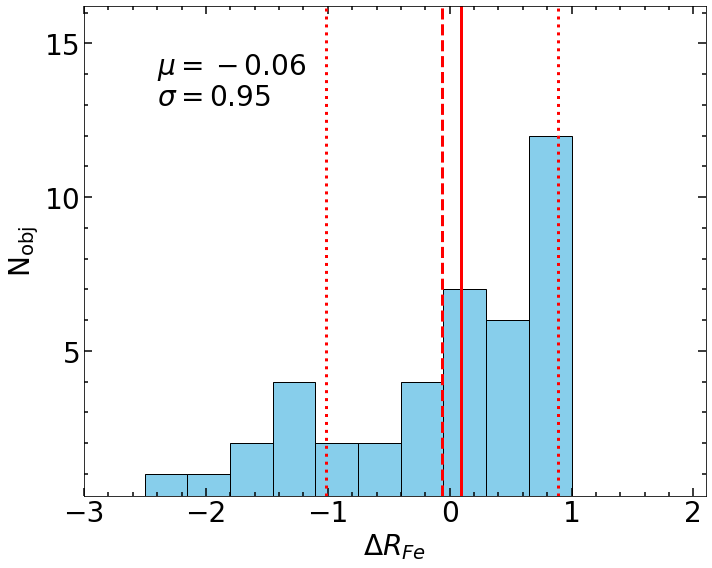}
    \caption{The $\Delta$\Rfe distribution. Although the fairly large spread, the mean relative difference is close to zero, thus our estimates are not systematically offset with respect to the SDSS control sample.}
    \label{fig:feii_check}
\end{figure}

Here we explore the possible differences in the \Rfe measurements between our low-luminosity AGN and the SDSS ones, whose properties are described in \citet{wu2022catalog}. To this aim, we compared the \Rfe values reported in the catalogue estimated through the \textsc{pyqsofit} code (\citealt{guo2018pyqsofit}) to those measured via our spectral fitting code, adopting the same Gaussian line deconvolution for the \Hb complex adopted in the catalogue. The most noticeable difference here is that the \textsc{pyqsofit} code employs the \citet{boroson1992emission} \Feii template, while we rely on \textsc{cloudy} models of \Feii emission. To ensure a meaningful test sample with data of decent quality, we performed the following selection process. We started from the control sample ($\sim$26,600 sources), and selected all the objects with low \Rfe ($\leq$0.15) and \Rfe/$\sigma_{\rm R_{Fe}}\geq$3. We sorted these sources by their average continuum SNR and picked the top 50. We then performed a spectral fit on these sources using our fitting code. In Fig.\ref{fig:feii_check} we show the distribution of the residuals defined as $\Delta$\Rfe$=$(\Rfe$_{\rm SDSS}$-\Rfe$_{\rm ours})$/\Rfe$_{\rm SDSS}$. Although there is quite a significant spread (the standard deviation approaches 100\%), we note that the mean is close to zero ($\langle \Delta$\Rfe$\rangle$=-0.04). This highlights that our fitting technique is not biasing on average the \Rfe measurements with respect to the control sample. Additionally, the tail of negative values, implies that in the cases of highest inconsistency our measurements tend to overestimate \Rfe with respect to the SDSS. Therefore, in these cases, the adoption of the alternative recipe for evaluating \Rfe, would result in even lower \Rfe values, thus going in the direction of even strengthening the observed difference.

%%%%%%%%%%%%%%%%%%%%%%%%%%%%%%%%%%%%%%%%%%%%%%%%%%%%%%%%%%%%%%%
% LOG DELTA RFe
%%%%%%%%%%%%%%%%%%%%%%%%%%%%%%%%%%%%%%%%%%%%%%%%%%%%%%%%%%%%%%%

\section{Spectral fits of the composite spectra}
\label{app:lowL_spec_fits}
\begin{figure}[h!]
	\includegraphics[width=\columnwidth]{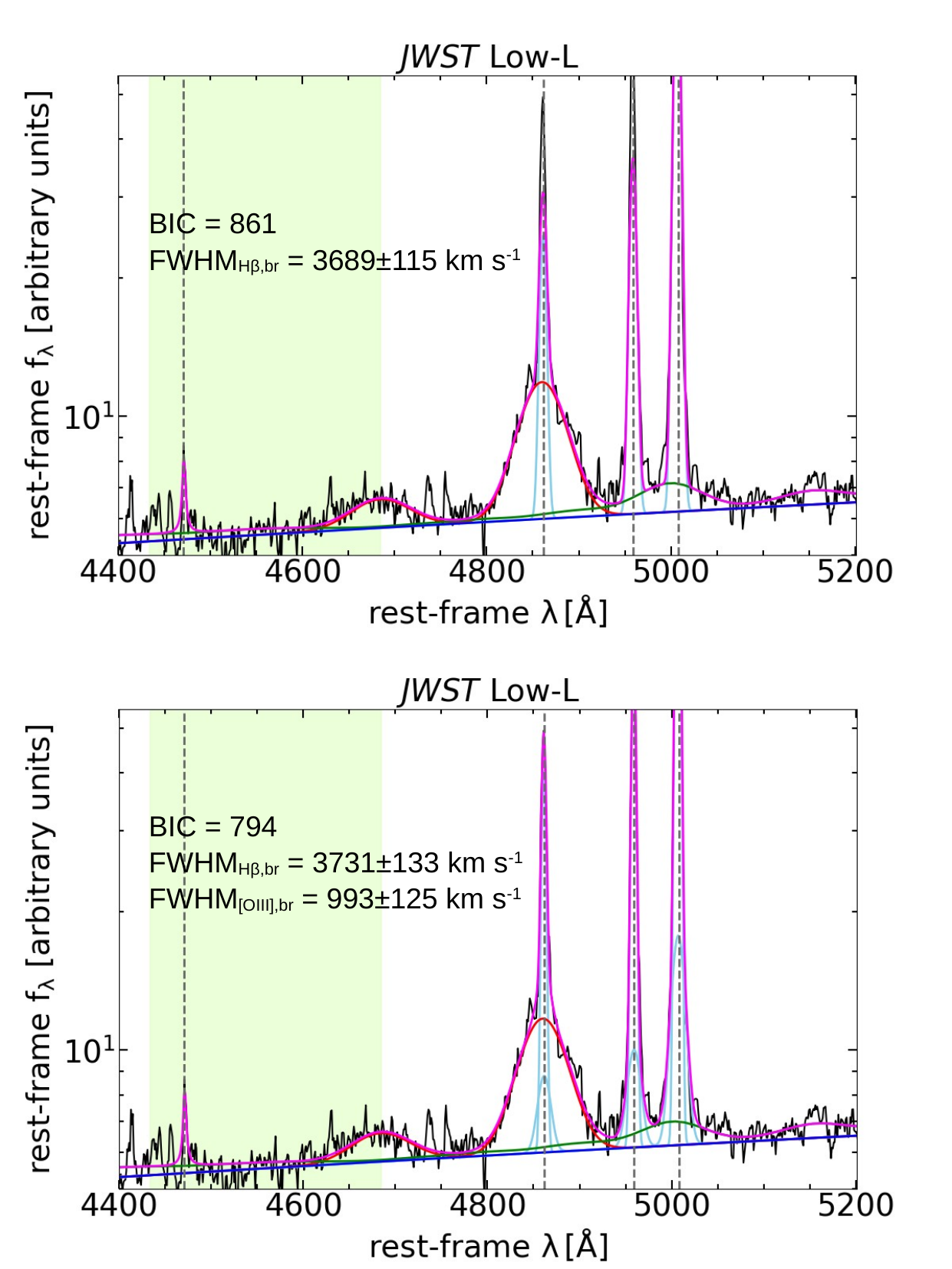}
    \caption{The spectral modelling of the \Hb region of the low-luminosity composite spectrum. The colour-coding is the same as in Fig.\ref{fig:fits_atlas}. In the top panel we show the fit where only one NLR narrow component is included. In the bottom panel a narrow and a broad outflow component are included for the NLR. The improvement in the best fit BIC is significant ($\Delta$BIC=-67). The $\rm FWHM$ of the broad [\ion{O}{iii}] (azure) is significantly smaller than that of the broad \Hb supposedly tracing the BLR (red).}
    \label{fig:lowL_stack_o3_fit}
\end{figure}

\begin{figure*}[h!]
	\includegraphics[width = \textwidth]{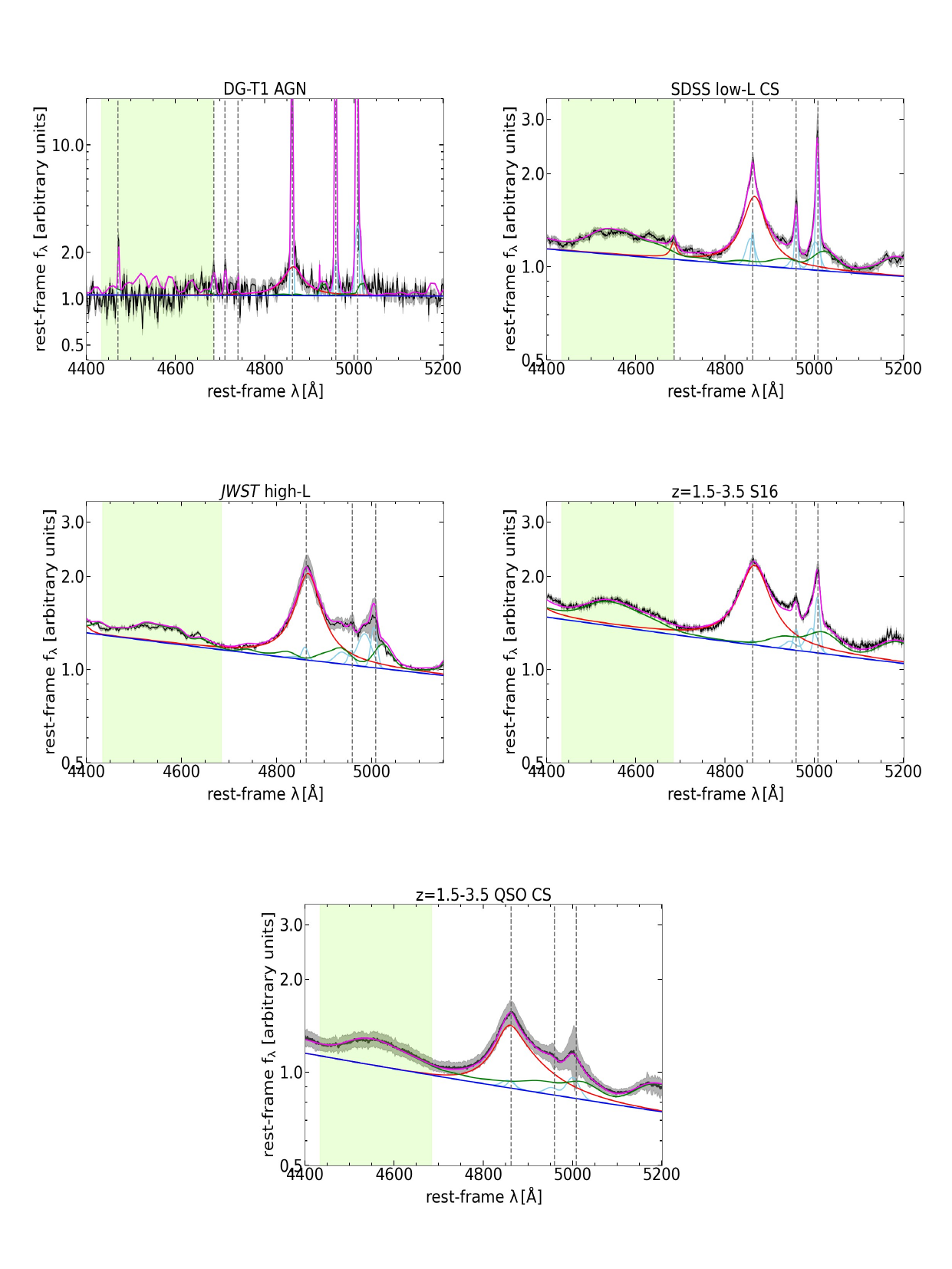}
    \caption{The spectral modelling of the \Hb--[\ion{O}{iii}] region for the composite spectra shown in Fig.\ref{fig:composites}. The colour-coding is the same as in Fig.\ref{fig:fit_ex}.}
    \label{fig:composites_fits}
\end{figure*}

In this Section we show the spectral fits of the composite spectra shown in Fig. \ref{fig:composites}. The colour-coding is the same as in Fig. \ref{fig:fit_ex}. In particular, in Fig. \ref{fig:lowL_stack_o3_fit}, we present the fit of the low-luminosity composite spectrum. In the top panel we show the fit without the broad component for the [\ion{O}{iii}] doublet, while in the bottom figure it is included. There, we also report the values of the broad \Hb (red) and [\ion{O}{iii}] (azure) $\rm FWHM$ values, which are significantly different. In Fig. \ref{fig:composites_fits} we display also the spectral fits of the other composite spectra built employing the control samples as well as the AGN in metal-poor dwarf galaxies.

%%%%%%%%%%%%%%%%%%%%%%%%%%%%%%%%%%%%%%%%%%%%%%%%%%%%%%%%%%%%%%%
% low-L O3 stack fit
%%%%%%%%%%%%%%%%%%%%%%%%%%%%%%%%%%%%%%%%%%%%%%%%%%%%%%%%%%%%%%%
\newpage

%%%%%%%%%%%%%%%%%%%%%%%%%%%%%%%%%%%%%%%%%%%%%%%%%%%%%%%%%%%%%%%
% low-L O3 stack fit

%%%%%%%%%%%%%%%%%%%%%%%%%%%%%%%%%%%%%%%%%%%%%%%%%%%%%%%%%%%%%%%

\newpage
\section{Fits atlas}
\label{app:fits_atlas}

Here we present the complete atlas of the spectral fits described in \ref{sec:spectral_fits}, together with the rest-frame spectra. Lines produced in the BLR are highlighted in red, those arising from the NLR in light blue, the \Feii is marked in green, while the power-law continuum is shown in blue. The various component are labelled as described in the top left plot. We also highlight in green the region where we evaluated the \Feii EW.

%%%%%%%%%%%%%%%%%%%%%%%%%%%%%%%%%%%%%%%%%%%%%%%%%%%%%%%%%%%%%%%
% FITS ATLAS 1
\begin{figure*}
	\includegraphics[width=\textwidth]{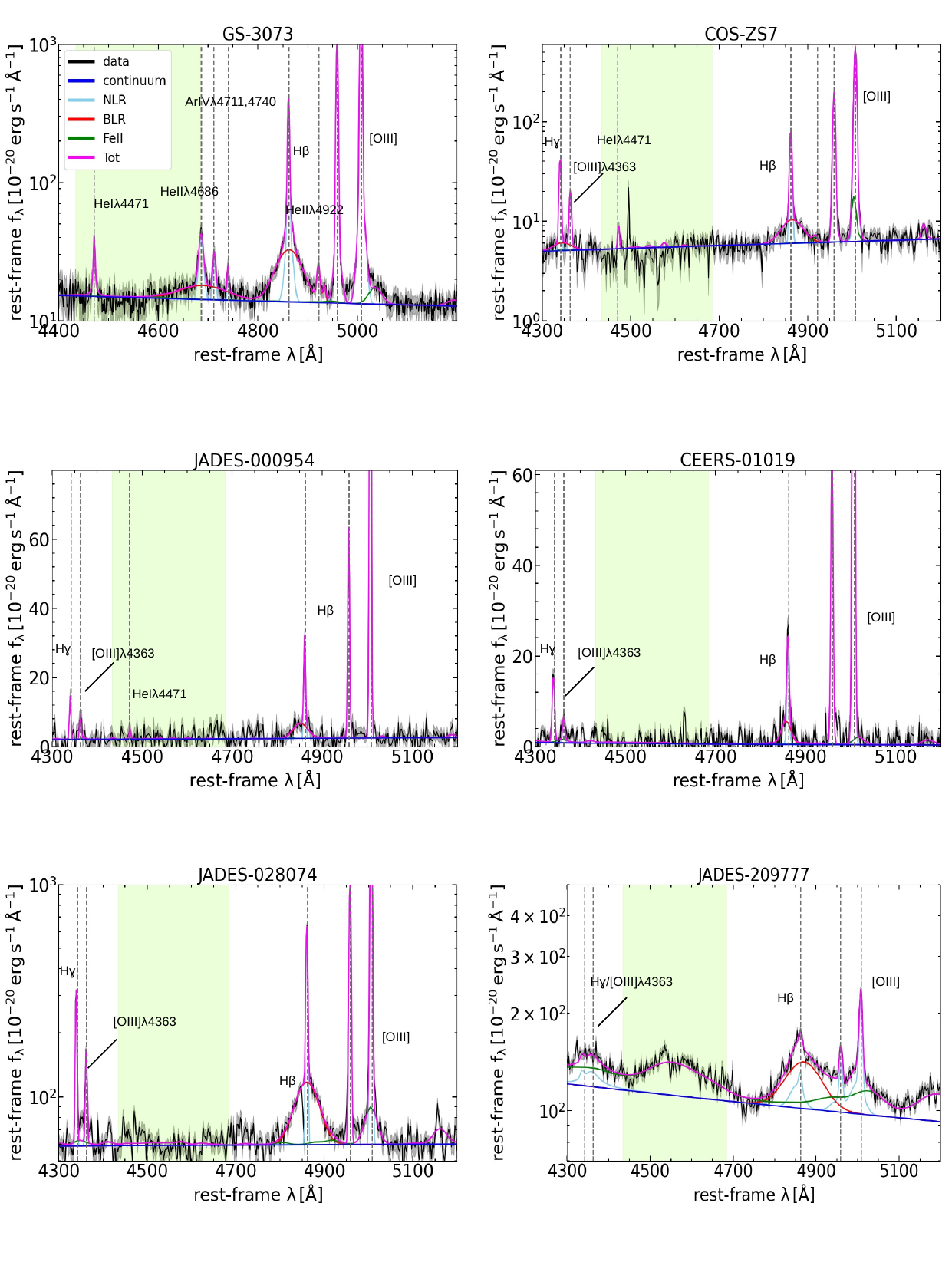}
    \caption{Fits atlas.}
    \label{fig:fits_atlas}
\end{figure*}
%%%%%%%%%%%%%%%%%%%%%%%%%%%%%%%%%%%%%%%%%%%%%%%%%%%%%%%%%%%%%%%
\newpage
%%%%%%%%%%%%%%%%%%%%%%%%%%%%%%%%%%%%%%%%%%%%%%%%%%%%%%%%%%%%%%%
% FITS ATLAS 2
\addtocounter{figure}{-1}
\begin{figure*}
	\includegraphics[width=\textwidth]{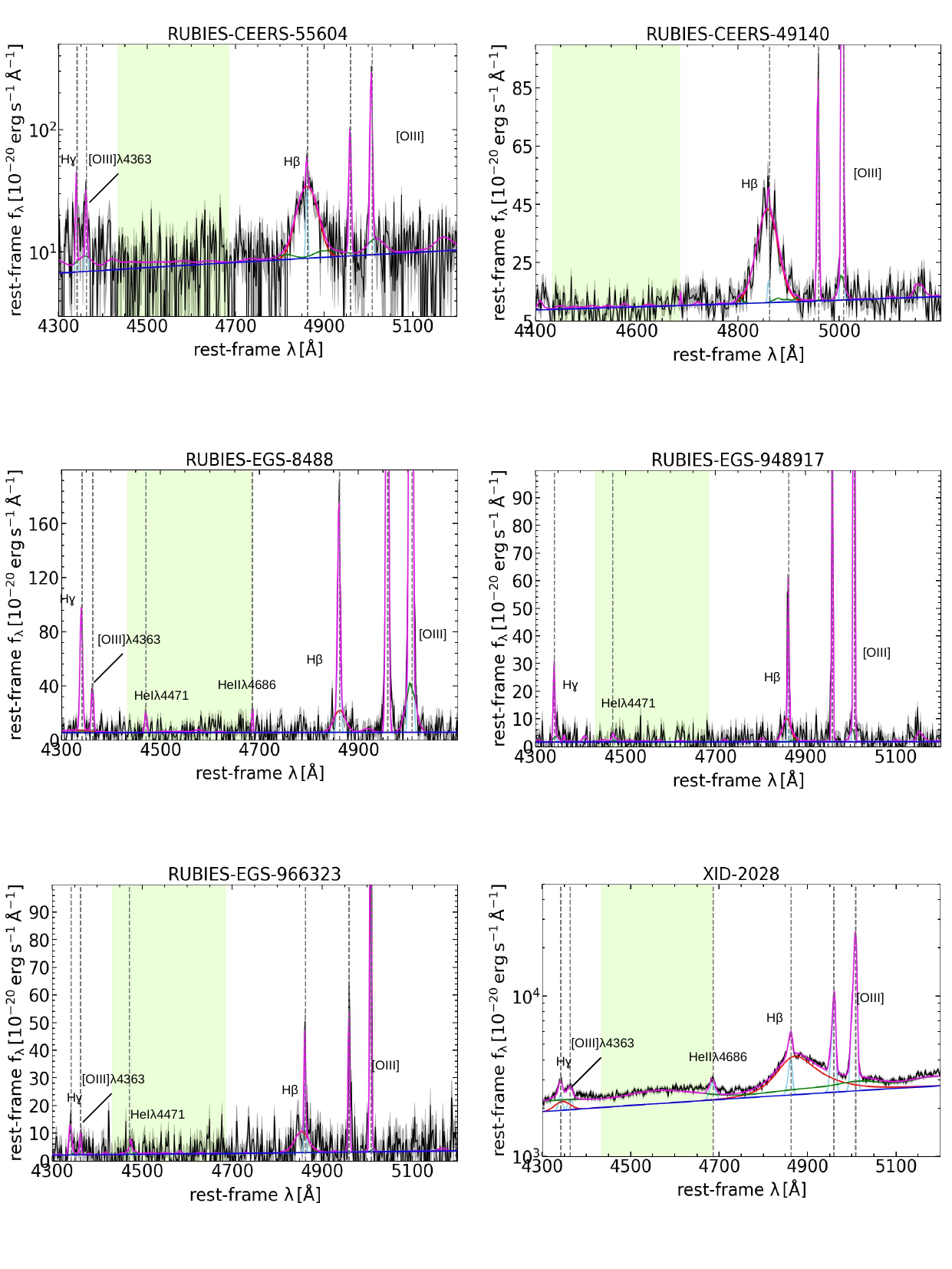}
    \caption{Fits atlas, continued.}
\end{figure*}
%%%%%%%%%%%%%%%%%%%%%%%%%%%%%%%%%%%%%%%%%%%%%%%%%%%%%%%%%%%%%%%
\newpage
% FITS ATLAS 3
\addtocounter{figure}{-1}
\begin{figure*}
	\includegraphics[width=\textwidth]{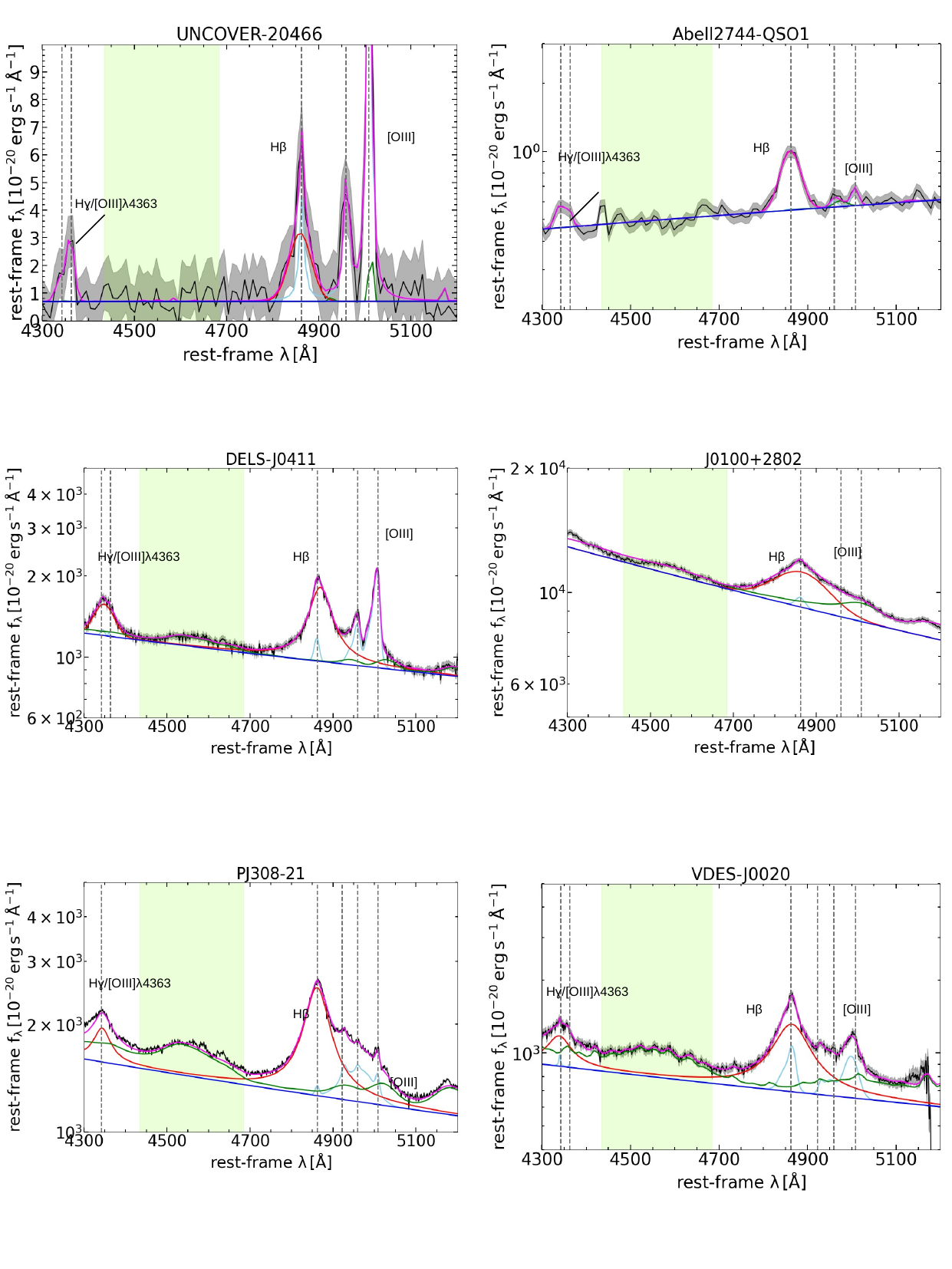}
    \caption{Fits atlas, continued.}
\end{figure*}

%%%%%%%%%%%%%%%%%%%%%%%%%%%%%%%%%%%%%%%%%%%%%%%%%%%%%%%%%%%%%%%
\newpage
% FITS ATLAS 3
\addtocounter{figure}{-1}
\begin{figure*}
	\includegraphics[width=\textwidth]{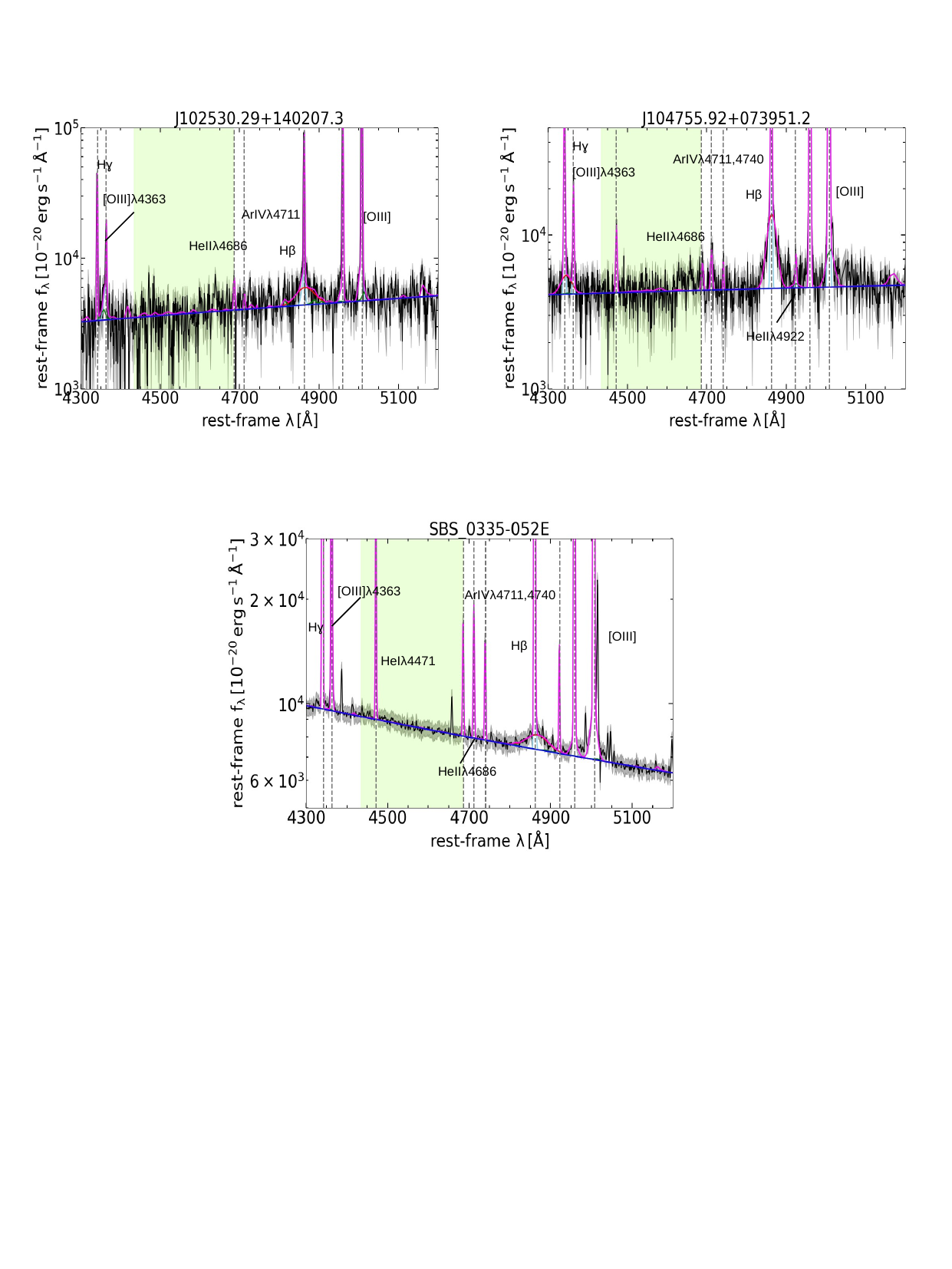}
    \caption{Fits atlas, continued.}
\end{figure*}
\newpage
%%%%%%%%%%%%%%%%%%%%%%%%%%%%%%%%%%%%%%%%%%%%%%%%%%%%%%%%%%%%%%%

\section{Photoionisation models with varying densities}
\label{appendix:varyden}

\begin{figure*}
    \centering
    \includegraphics[width=\textwidth]{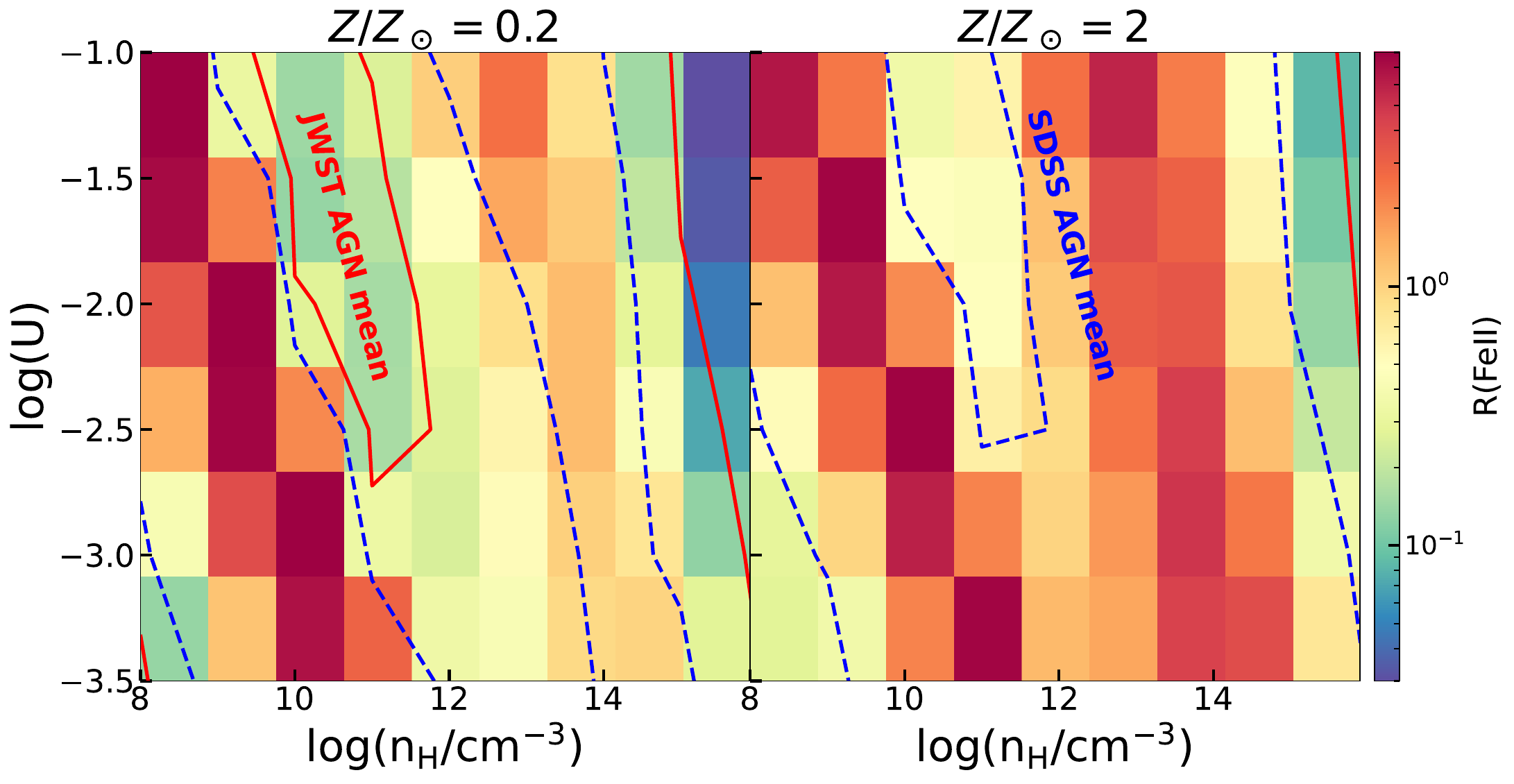}
    \caption{Photoionisation model predicted \Rfe as a function of the ionisation parameter and hydrogen density. \textit{Left:} Models with a subsolar metallicity of $Z/Z_\odot = 0.2$ plausible for BLRs of low-luminosity AGN observed by \textit{JWST}. \textit{Right:} Models with a super solar metallicity of $Z/Z_\odot = 2$ plausible for BLRs of local AGN as well as high-luminosity quasars observed in SDSS. In both figures, the mean \Rfe of \textit{JWST}-identified AGN is indicated by the solid red contour, and the mean \Rfe of SDSS AGN is indicated by the dashed blue contour.
    }
    \label{fig:model_varyden}
\end{figure*}

In this Appendix we present additional photoionisation models with a range of hydrogen densities from $10^8~{\rm cm^{-3}}$ to $10^{16}~{\rm cm^{-3}}$. In Figure~\ref{fig:model_varyden} we show two sets of AGN BLR models with subsolar and supersolar metallcities. Apart from the hydrogen density and the metallicity, all other parameters are set to be the same as described in Table~\ref{tab:models}. It is clear that for both sets of models, local maxima of \Rfe are found for certain combinations of $U$ and $n_{\rm H}$, a phenomenon already noted by previous modelling works \citep[e.g.,][]{baldwin2004origin,ferland_fe2_2009}.

As a comparison, we also show the contours corresponding to the means of JWST-identified Type 1 AGN and SDSS Type 1 AGN with similar $\rm FWHM_{\Hb,br}$ and $\rm L_{\Hb,br}$, respectively. Compared to the subsolar metallicity models ($Z=0.2~Z_\odot$), the \textit{JWST} AGN mean is found for $10^{10}~{\rm cm^{-3}}<n_{\rm H}<10^{12}~{\rm cm^{-3}}$. In contrast, the supersolar metallicity models ($Z=2~Z_\odot$) only match the \textit{JWST} AGN mean when $n_{\rm H} \gtrsim 10^{16}~{\rm cm^{-3}}$. The SDSS mean can be found at $10^{10}~{\rm cm^{-3}}<n_{\rm H}<10^{12}~{\rm cm^{-3}}$,  $n_{\rm H} \lesssim 10^{9}~{\rm cm^{-3}}$, or $n_{\rm H} \gtrsim 10^{15}~{\rm cm^{-3}}$ compared to the supersolar metallicity models depending on the value of $U$, or a broader density range compared to the subsolar metallicity models.

While Figure~\ref{fig:model_varyden} illustrates the complex degeneracies between model parameters, as we have discussed in Sec.~\ref{sec:cloudy}, based on the current observational evidence it might be more natural to assume redshift evolution in the metallicity rather than other model parameters for BLRs.
A scenario invoking the redshift evolution in the gas density, for example, would require ultra-dense environments for JWST-identified AGN assuming they are similarly metal enriched as SDSS AGN.
At extremely high densities, the strength of permitted UV lines from the BLR would change significantly \citep{temple2021high}, which could be tested observationally using medium-to-high resolution UV follow-ups of less obscured early AGN with \jwst.
However, this is beyond the scope of the current work and we leave it for future investigations.

\end{appendix}

\end{document}